\documentclass[runningheads]{llncs}

\usepackage{eccv}

\usepackage{eccvabbrv}
\usepackage{graphicx}
\usepackage{booktabs}
\usepackage{multirow}
\usepackage[table]{xcolor}
\usepackage{makecell}
\usepackage{subcaption}
\usepackage{algorithm}
\usepackage{algorithmic}
\usepackage{amsmath}
\usepackage{amssymb}
\usepackage[accsupp]{axessibility}
\usepackage{placeins}
\usepackage[hidelinks]{hyperref}
\usepackage{orcidlink}

\begin{document}

\title{NGPS: Structure-Preserving Self-Supervised Denoising via
Neighbor-Guided Patch Sampling}

\titlerunning{NGPS for Self-Supervised Denoising}

\author{
Jaehyun Cho\inst{1}\orcidlink{0009-0007-9135-1509}
\and
YoungJoon Yoo\inst{1,2}\orcidlink{0009-0000-8778-5573}
\thanks{Corresponding author.}
}
\authorrunning{J.~Cho and Y.~Yoo}
\institute{
Department of Artificial Intelligence,
Chung-Ang University,
84 Heukseok-ro, Dongjak-gu, Seoul, Republic of Korea\\
\email{\{q7011805,yjyoo3312\}@cau.ac.kr}
\and
SNUAILAB,
1 Gwanak-ro, Gwanak-gu, Seoul 08826, Korea\\
\email{yjyoo3312@snuailab.ai}
}

\maketitle

\begin{abstract}
Neighboring-slice self-supervised denoising is attractive for volumetric medical imaging, yet inter-slice misalignment breaks anatomical correspondence and often yields ghosting and blurred margins when adjacent slices are used na\"ively as targets.
We propose Neighbor-Guided Patch Sampling (NGPS), a lightweight framework that constructs neighboring supervision under local inter-slice misalignment.
To avoid learning from misleading targets, prior methods commonly mask discrepant regions, but this stabilizes training at the cost of leaving a non-trivial portion of neighboring evidence unexploited, particularly around high-frequency anatomical boundaries.
NGPS addresses this by \emph{decoupling} structure matching from signal retrieval: for each masked location, it searches a local neighborhood for structurally similar candidate patches using a simple guide image (\eg, fast bilateral filtering), while retrieving the supervision signal directly from the \emph{raw noisy} neighbor at the matched coordinates. By matching on a noise-attenuated guide while retrieving raw values from neighboring slices, NGPS constructs local pseudo targets without dense deformation-field estimation or spatial resampling.
Across the evaluated CT and synthetic-Rician MRI settings, NGPS improves fidelity and structure-sensitive metrics. Code is available at \url{https://github.com/cv-cho/NGPS}.
\end{abstract}

\section{Introduction}
\label{sec:intro}

\begin{figure}[t]
\centering

\begin{subfigure}{\textwidth}
    \centering
    \includegraphics[width=\textwidth]{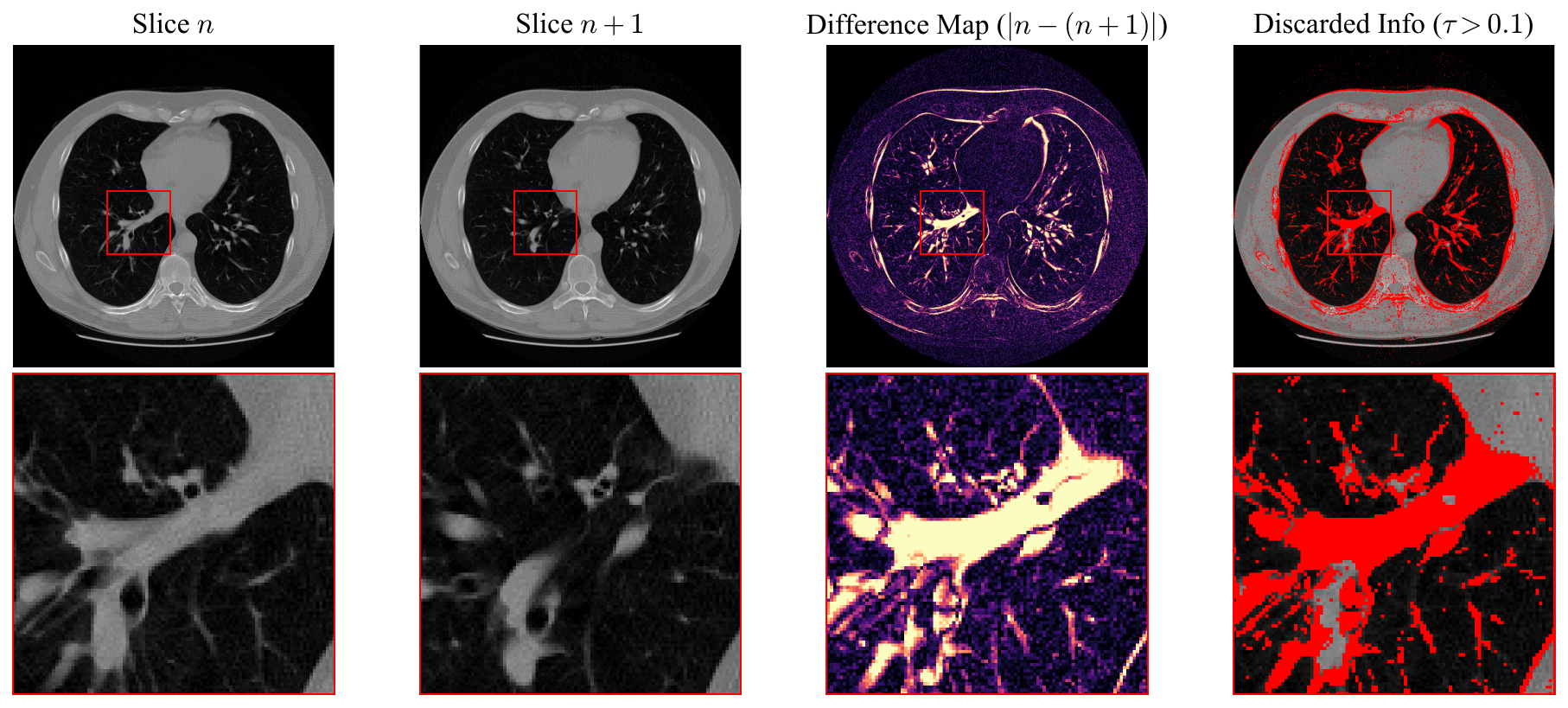}
    \caption{Qualitative visualization of discrepancy-based masking}
    \label{fig:misalignment_a}
\end{subfigure}
\begin{subfigure}[t]{0.48\textwidth}
    \centering
    \includegraphics[width=\linewidth]{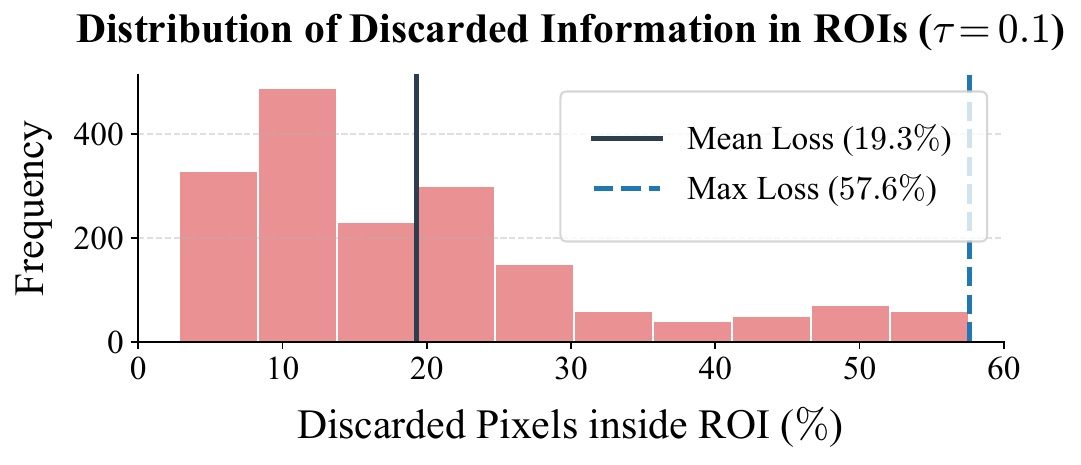}
    \caption{Distribution of discarded information}
    \label{fig:misalignment_b}
\end{subfigure}
\hfill
\begin{subfigure}[t]{0.48\textwidth}
    \centering
    \includegraphics[width=\linewidth]{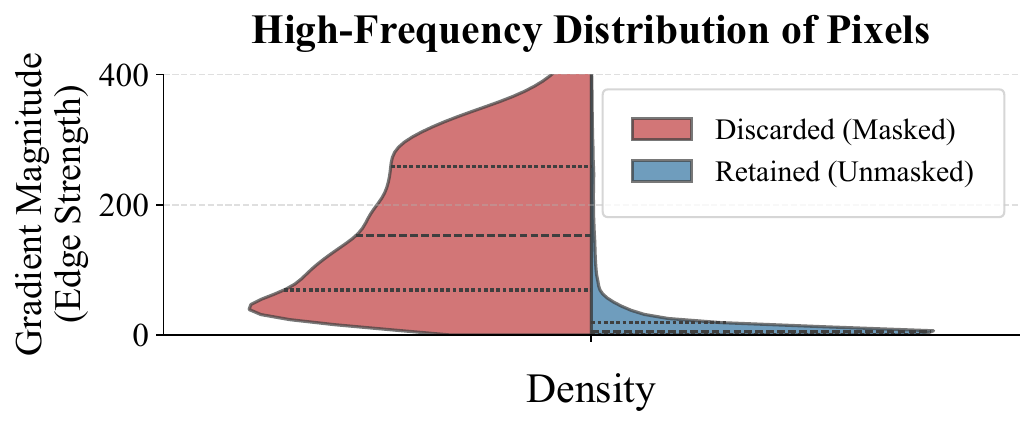}
    \caption{High-frequency distribution of pixels}
    \label{fig:misalignment_c}
\end{subfigure}

\caption{Illustration of inter-slice anatomical misalignment and the resulting critical information loss from conventional masking strategies ($\tau = 0.1$) in 2.5 mm LIDC-IDRI data \cite{Armato_2011_LIDC}. 
\textbf{(a)} The absolute difference map ($|n - (n+1)|$) and the masked pixels (highlighted in red).
\textbf{(b)} The histogram revealing the severity of this spatial information loss within critical Regions of Interest (ROIs) across the validation set.
\textbf{(c)} The split violin plot showing that the discarded (masked) pixels are heavily concentrated in the high-frequency spectrum (high gradient magnitude) compared to the retained pixels.
}
\label{fig:misalignment}
\end{figure}

Medical imaging enables non-invasive diagnosis and longitudinal monitoring \cite{Sagheer_2020_Review}. 
However, X-ray-based modalities such as CT and fluoroscopy expose patients and clinicians to ionizing radiation, motivating low-dose acquisition \cite{Li_2020_SACNN}. 
Dose reduction induces photon starvation, yielding quantum noise and reconstruction artifacts that obscure fine anatomical structures, especially in high-attenuation regions \cite{Andreozzi_2020_Index,Cesarelli_2013_Noise,McCollough_2017_AAPM}. 
This motivates denoising methods that restore structural and signal fidelity while preserving clinically meaningful content \cite{Luo_2020_Ultra,Chen_2017_LowDose,Fourati_2005_Wavelet}.

Prior approaches include classical priors (\eg, NLM, BM3D, TV) \cite{Buades_2005_NLM,Manjon_2008_MRI,Dabov_2007_BM3D,Chambolle_2004_TV} and supervised deep networks \cite{Zhang_2017_DnCNN,Ronneberger_2015_UNet}, but paired clean targets are often infeasible and motion or anatomical variability further complicates alignment \cite{McCollough_2017_AAPM,Lee_2017_Motion,Weigert_2018_CARE}. 
Self-supervised learning (SSL) alleviates this constraint by learning directly from noisy observations \cite{Lehtinen_2018_N2N,Krull_2019_N2V,Batson_2019_N2S,Xie_2020_Noise2Same}. 
For volumetric data, neighboring-slice SSL uses adjacent slices as supervision \cite{Zhou_2024_NSN2N, Niu_2023_Noise2Sim, Xu_2021_Deformed2Self}, yet inter-slice misalignment breaks the same-coordinate correspondence, so relevant anatomical evidence is often \emph{displaced} rather than missing. Fig.~\ref{fig:misalignment} highlights that discrepancy-based masking can exclude a substantial portion of pixels within clinically relevant ROIs, removing \emph{on average about 20\%} of the available neighboring supervision. 
Moreover, the excluded regions are concentrated around high-gradient anatomical boundaries.
This creates a persistent dilemma: masking-based remedies stabilize training by avoiding misleading targets, but systematically withhold supervision in structurally challenging regions. 
Registration-based alternatives explicitly estimate correspondences and resample neighboring slices. Under severe noise or larger inter-slice gaps, the estimated alignment can become less reliable, while spatial resampling may smooth edges and add alignment overhead. These trade-offs motivate local target retrieval without dense warping.
Therefore, a key challenge is to retrieve displaced anatomical evidence from a local neighborhood rather than treating it as unavailable.

Motivated by this, we propose \emph{Neighbor-Guided Patch Sampling (NGPS)}, a misalignment-aware neighboring-slice SSL framework that is designed to (i) recover displaced supervision within a local neighborhood without dense deformation-field estimation or spatial warping, (ii) remain compatible with Noise2Noise-style \cite{Lehtinen_2018_N2N} training, and (iii) preserve high-frequency anatomical margins with minimal overhead. 
NGPS reframes misalignment handling as a \emph{training-time supervision construction} problem, rather than an inference-time non-local denoising prior. 
Specifically, NGPS constructs a neighboring pixel bank by matching pre-filtered guide patches across adjacent slices and retrieving target values from the corresponding raw noisy neighbors. The primary contributions are:
\begin{itemize}
\item \textbf{Misalignment as displaced supervision:} We identify a structural supervision gap in neighboring-slice SSL caused by displacement, clarifying the masking versus registration trade-off.
\item \textbf{Neighbor-Guided Patch Sampling (NGPS):} We introduce a lightweight patch-based supervision recovery mechanism using guide-feature search and raw-target retrieval for misalignment-aware SSL.
\item \textbf{Comprehensive validation:} We demonstrate consistent gains across low-dose CT and MRI benchmarks (AAPM-Mayo \cite{McCollough_2017_AAPM}, LIDC-IDRI \cite{Armato_2011_LIDC}, IXI \cite{Biomedical_2018_IXI}), with improved preservation of fine anatomical margins.
\end{itemize}

\section{Related Work}

\paragraph{Self-supervised Learning for Denoising.}
To overcome the difficulty of obtaining paired noisy-clean images \cite{McCollough_2017_AAPM,Weigert_2018_CARE}, self-supervised learning (SSL) has become a practical alternative. Early image-domain methods, such as Blind-Spot Networks \cite{Krull_2019_N2V,Batson_2019_N2S}, construct supervision without directly observing the target pixel, typically through masking or blind-spot/downsampling designs. While effective in broad settings, these schemes restrict the directly available context and can introduce smoothing or checkerboard artifacts \cite{Wang_2022_B2U}. 
Recent image-domain SSL methods broaden target construction through masking, downsampling, noise injection, or generative pseudo targets. DiffDenoise \cite{Demir_2025_DiffDenoise}, for example, uses a conditional-diffusion pipeline to generate pseudo-clean supervision. These methods differ in how supervision is synthesized or withheld. In the neighboring-slice setting studied here, Fig.~\ref{fig:misalignment} shows that discrepancy masking removes supervision concentrated near high-gradient anatomical boundaries.

\paragraph{Volumetric SSL and Misalignment Handling.}
Volumetric medical SSL can exploit redundancy along different axes. Patch2Self \cite{Fadnavis_2020_Patch2Self} exploits diffusion-MRI q-space redundancy through held-out-volume regression. Our setting instead uses spatial redundancy across adjacent anatomical slices, which provide alternative noisy observations under the standard slice-independent-noise assumption. This neighboring-slice formulation implicitly assumes anatomical correspondence; in practice, inter-slice displacement can make same-coordinate supervision blur shifted structures. Existing remedies largely follow registration- or masking-based strategies.
Registration-based methods such as MSR2AU-Net \cite{Jeon_2025_MSR2AUNet} and Deformed2Self \cite{Xu_2021_Deformed2Self} estimate correspondences and warp neighboring slices. Such alignment can become less reliable under severe degradation or larger inter-slice gaps, and spatial resampling may alter local signal and noise statistics.
Masking-based methods such as Noise2Sim \cite{Niu_2023_Noise2Sim} and NS-N2N \cite{Zhou_2024_NSN2N} instead exclude discrepant pixels from the reconstruction loss. This avoids misleading same-coordinate targets but can reduce supervision near high-gradient boundaries, as quantified in Fig.~\ref{fig:misalignment}. 
NGPS addresses this displaced-supervision regime through local retrieval rather than dense warping or exclusion.

\paragraph{Patch-based Sampling and Neighbor-Aware Processing.} 
Patch-based self-similarity provides a complementary direction for constructing supervision without paired targets. Pixel2Pixel (P2P) \cite{Ma_2025_Pixel2Pixel} constructs a ``PixelBank'' by searching for similar patches within a single noisy image and sampling replacement targets from non-local neighbors. Extending this idea to volumetric SSL is non-trivial: exhaustive image-wide search can be costly, and same-image target selection can couple the retrieval process to the same noisy realization. Neighboring slices provide an alternative target source under the standard slice-independent-noise assumption, provided displaced correspondence can be resolved. 
At a broader conceptual level, Graph Flow Matching \cite{Siddiqui_2026_GraphFlow} introduces neighbor-aware aggregation in a flow-based image-generation setting rather than for denoising-target construction. NGPS differs from both settings by applying local guide-based patch search at locations flagged by inter-slice discrepancy and retrieving raw adjacent-slice values as training targets.

\section{Methodology}
\label{sec:method}

\begin{figure*}[t]
\centering
\includegraphics[width=\textwidth]{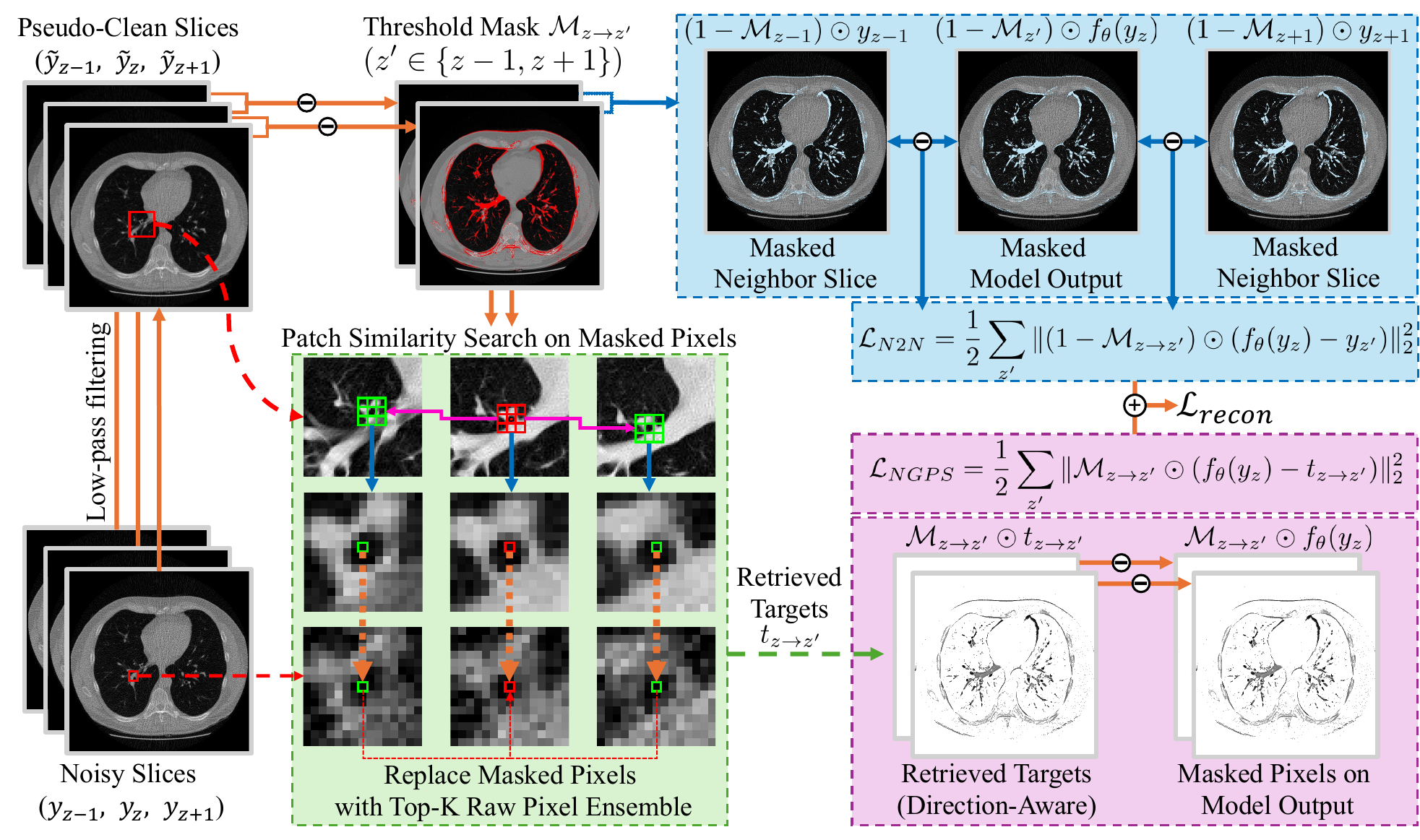}
\caption{\textbf{Overall NGPS pipeline.}
A noisy slice triplet is low-pass filtered to produce noise-attenuated guide (``Pseudo-Clean'') slices, whose pairwise differences yield direction-aware threshold masks with reduced sensitivity to raw noise.
For each masked location, NGPS searches a local window in the adjacent slice for the Top-$K$ similar patches (green); the matched coordinates are then used to retrieve and average the corresponding center-pixel values from the original noisy adjacent slice, forming direction-aware retrieved targets.
Unmasked static pixels use same-coordinate neighboring supervision to compute $\mathcal{L}_{N2N}$ (blue), whereas masked pixels use the retrieved targets to compute $\mathcal{L}_{NGPS}$ (purple).
The two reconstruction terms are summed as $\mathcal{L}_{recon}$; regional consistency is applied separately to static regions as described in Sec.~\ref{subsec:objective}.}
\label{fig:method}
\end{figure*}

We propose Neighbor-Guided Patch Sampling (NGPS), a misalignment-aware neighboring-slice SSL framework that constructs pseudo targets for misaligned regions via three steps, as illustrated in Fig.~\ref{fig:method}: (i) fast guide generation with conservative masking, (ii) decoupled structural matching on the guide, and (iii) raw target retrieval with a Top-$K$ ensemble. We then train a 2D slice denoiser $f_\theta$ with a hybrid objective that uses standard same-coordinate supervision in static regions (blue in Fig.~\ref{fig:method}) and switches to NGPS-recovered targets in misaligned regions (purple). Furthermore, we incorporate an output-level regional consistency regularizer to reduce slice-to-slice flicker in non-masked areas.

\subsection{Problem Setup and Motivation}
Let $x, y \in \mathbb{R}^{H \times W \times D}$ denote a clean volumetric image and its noisy observation, modeled as $y_z(p) = x_z(p) + n_z(p)$, where $z\in\{1,\ldots,D\}$ is the slice index and $p$ is the spatial coordinate. Assuming zero-mean ($\mathbb{E}[n_z(p)]=0$) and slice-independent noise, neighboring-slice SSL learns a denoiser using adjacent slices as pseudo targets \cite{Lehtinen_2018_N2N}. Here, slice independence concerns noise across the input and target slices; it does not require pixel-wise i.i.d.\ noise within each slice.
However, inter-slice misalignment breaks this same-coordinate correspondence. Under an inter-slice displacement $\delta$, a local anatomical structure in the clean signal $x_z(p)$ appears shifted in the adjacent slice as $x_{z'}(p) \approx x_z(p-\delta)$. Naive same-coordinate supervision can encourage averaging between these shifted structures, yielding a superposition $\frac{1}{2}\big(x_z(p)+x_z(p-\delta)\big)$ that can produce ghosting and blurred anatomical boundaries.

\subsection{From Misalignment to Design Principles}
\label{subsec:design_principles}
To overcome this dilemma, NGPS reframes misalignment as \emph{local displacement} rather than absence. Instead of discarding misaligned regions, we actively retrieve the corresponding anatomical evidence from a nearby coordinate $q$ in the adjacent slice. This perspective establishes three core design requirements for our framework:
(i) \textbf{Recover displaced supervision} via explicit local retrieval rather than passive exclusion,
(ii) \textbf{Retain raw neighboring values as training targets} while
using filtered guides only for correspondence search, and
(iii) \textbf{Remain lightweight} without dense deformation-field estimation or spatial resampling.

We use \emph{selection bias} to denote noise-driven coordinate selection that induces correlation between input noise and the residual noise of a retrieved target. NGPS performs discrete local correspondence search. Unlike dense registration, it estimates no dense deformation field and performs no spatial warping or interpolation of neighboring slices.

\subsection{Neighbor-Guided Patch Sampling (NGPS)}
\label{subsec:ngps}
For each pixel flagged as misaligned, NGPS constructs a pseudo target $t_{z \to z'}(p)$ via
(i) fast guide generation with conservative masking,
(ii) decoupled patch matching on the guide,
and (iii) raw-value retrieval with Top-$K$ aggregation.

\subsubsection{Guide generation and misalignment mask.}
We generate an edge-preserving guide volume $\tilde{y}$ by sequentially applying a 2D bilateral filter (BF) and a median filter (MF) to suppress quantum noise while preserving structural boundaries:

\begin{equation}
\tilde{y}_z(p) = \mathrm{MF}\Big(\mathrm{BF}\big(y_z(p); \sigma_s, \sigma_r\big); \kappa\Big),
\end{equation}

where $\sigma_s$ and $\sigma_r$ are the spatial and range parameters of the BF, and $\kappa$ is the kernel size of the MF.
Let $\mathcal{N}(z)=\{z-1,z+1\}$ denote the set of available neighboring slice indices (applying reflection padding only at the volume ends). Instead of a single global mask, we define a direction-aware misalignment mask for each neighbor $z' \in \mathcal{N}(z)$ based on the same-coordinate guide disagreement:

\begin{equation}
\mathcal{M}_{z \to z'}(p) =
\mathbb{1}\Big(\big|\tilde{y}_z(p)-\tilde{y}_{z'}(p)\big| > \tau \Big),
\end{equation}

where $\tau$ is used to flag locations whose same-coordinate guide discrepancy exceeds the selected threshold. For pixels not flagged by $\mathcal{M}_{z\to z'}$, we use standard same-coordinate neighboring supervision. For flagged pixels, we avoid same-coordinate targets and invoke local NGPS retrieval.

\subsubsection{Decoupled matching and raw retrieval.}
For each pixel $p \in \mathbb{Z}^2$ flagged as misaligned with respect to a specific neighboring slice $z' \in \mathcal{N}(z)$, we search for structurally similar candidate patches within a local spatial window $\Omega_p$ (\eg, $15{\times}15$) in slice $z'$. 
While the similarity search is performed at the patch level to ensure structural context, the supervision signal is retrieved purely as a single scalar value from the center pixel of the matched location.
Let $\mathcal{P}_k(I,p) \in \mathbb{R}^{k \times k}$ extract a $k\times k$ patch centered at $p$ from image $I$.
We compute the guide-based sum of squared differences (SSD) costs:
\begin{equation}
\mathcal{D}(p,q;z') = \left\| \mathcal{P}_k(\tilde{y}_z,p) - \mathcal{P}_k(\tilde{y}_{z'},q) \right\|_2^2,
\quad q\in\Omega_p,
\end{equation}
and subsequently retrieve the candidate target scalar from the \emph{raw} center pixel, $y_{z'}(q) \in \mathbb{R}$.
This separates correspondence search from regression-target retrieval, avoiding direct raw-patch matching while keeping the target value unfiltered.

The guide is computed from noisy observations and therefore does not make the selected coordinates strictly independent of noise. We accordingly treat guide-based matching as an empirical design choice intended to reduce noise-driven target selection, while raw-value retrieval keeps the regression target unfiltered.
Supplementary Table~\ref{tab:supp_PixelBank_counterfactual} compares NGPS with the closest same-slice and adjacent-slice PixelBank-style alternatives.

\subsubsection{Top-$K$ ensemble target.}
To reduce the variance inherent to a single best match, we aggregate the Top-$K$ candidate pixels for each neighboring direction $z'$.
For a given neighbor $z'$, we rank the candidate coordinates in $\Omega_p$ and select the set
$\{q^{(1)},\ldots,q^{(K)}\}$ yielding the smallest $\mathcal{D}(p,q;z')$.
We define the direction-aware target $t_{z \to z'}(p)$ as the average of these $K$ retrieved center pixels:
\begin{equation}
t_{z \to z'}(p) = \frac{1}{K}\sum_{k=1}^{K} y_{z'}\big(q^{(k)}\big).
\end{equation}
This is inspired by the PixelBank philosophy \cite{Ma_2025_Pixel2Pixel} but specialized to volumetric SSL; the sampling is applied directionally (only for misaligned pixels) and the scalar targets are drawn exclusively from the adjacent slice $z'$ to construct $t_{z \to z'}$.

\subsection{Training objective}
\label{subsec:objective}

We optimize $f_\theta$ using a hybrid reconstruction loss and regional consistency term:
\begin{equation}
\mathcal{L}_{total} = \mathcal{L}_{recon} + \lambda\,\mathcal{L}_{RC},
\end{equation}
where $\lambda$ is a hyperparameter that controls the strength of the regularization.

\subsubsection{Hybrid reconstruction loss}
Guided by the directional misalignment mask $\mathcal{M}_{z \to z'}$, we adaptively switch the supervision signal. Specifically, we apply standard same-coordinate neighboring targets for static regions, and provide the retrieved NGPS targets $t_{z \to z'}$ for misaligned regions. The hybrid reconstruction loss is thus formulated as:
\begin{align}
\mathcal{L}_{recon}
&=
\mathcal{L}_{N2N}
+
\mathcal{L}_{NGPS},
\\
\mathcal{L}_{N2N}
&=
\frac{1}{|\mathcal N(z)|}
\sum_{z'\in\mathcal N(z)}
\frac{
\sum_p
\big(1-\mathcal M_{z\to z'}(p)\big)
\big(f_\theta(y_z)(p)-y_{z'}(p)\big)^2
}{
\sum_p
\big(1-\mathcal M_{z\to z'}(p)\big)
+\epsilon
},
\label{eq:n2n}
\\
\mathcal{L}_{NGPS}
&=
\frac{1}{|\mathcal N(z)|}
\sum_{z'\in\mathcal N(z)}
\frac{
\sum_p
\mathcal M_{z\to z'}(p)
\big(f_\theta(y_z)(p)-t_{z\to z'}(p)\big)^2
}{
\sum_p
\mathcal M_{z\to z'}(p)
+\epsilon
},
\end{align}
where $\mathcal{M}_{z \to z'}$ and $t_{z \to z'}$ are spatial maps matching the dimensions of the prediction $f_\theta(y_z)$, and $\epsilon>0$ is a small numerical constant preventing division by zero. Each directional loss is normalized by the number of pixels assigned to its corresponding supervision region.

\subsubsection{Regional consistency regularization}
Equation~\ref{eq:n2n} optimizes each slice independently and does not explicitly enforce volumetric coherence.
In regions identified as static, adjacent clean signals are expected to be locally similar rather than identical.
Following volumetric SSL practice \cite{Zhou_2024_NSN2N}, we therefore penalize prediction differences only on $(1-\mathcal{M}_{z \to z'})$, encouraging inter-slice consistency without constraining regions flagged as displaced:
\begin{equation}
\mathcal{L}_{RC} = \frac{1}{|\mathcal{N}(z)|}\sum_{z'\in\mathcal{N}(z)}
\big\| (1-\mathcal{M}_{z \to z'})\odot\big(f_\theta(y_z)-f_\theta(y_{z'})\big) \big\|_2^2.
\end{equation}

\section{Experiments}

\subsection{Experimental Setup}

\begin{table}[t]
\centering
\caption{Summary of datasets and noise simulation protocols used in our experiments.}
\label{tab:datasets}
\resizebox{\textwidth}{!}{%
\begin{tabular}{l|c|c|c|l|c}
\hline
\textbf{Dataset} & \textbf{Modality} & \textbf{Anatomy} & \textbf{Spacing} & \textbf{Noise Simulation Type} & \textbf{Train/Test} \\ \hline
AAPM-Mayo \cite{McCollough_2017_AAPM} & CT & Abdomen & 1.0 mm & Realistic Quarter-Dose (Poisson injection in sinogram) & 8 / 2 \\
LIDC-IDRI \cite{Armato_2011_LIDC} & CT & Thorax & 1.25, 2.5 mm & Simulated Ultra-Low-Dose (Radon Transform, $P=12.5K$) & 66 / 6 \\
IXI-T1 \cite{Biomedical_2018_IXI} & MRI & Brain & 1.2 mm & Synthetic Rician ($\sigma \in \{5\%, 7\%, 9\%\}$) & 52 / 5 \\ \hline
\end{tabular}%
}
\end{table}

\subsubsection{Datasets and Noise Simulation.}
To evaluate the proposed NGPS, we use three public medical volumetric datasets. The detailed configurations for each dataset are summarized in Table~\ref{tab:datasets}.

\paragraph{AAPM-Mayo (CT) \cite{McCollough_2017_AAPM}:} As a realistic benchmark for low-dose CT (LDCT) denoising, we use abdominal CT scans from the AAPM Low-Dose CT Grand Challenge. Unlike simple additive noise, the Quarter-Dose (QD) images were generated by injecting Poisson noise directly into the \textit{projection data} (sinograms) of Normal-Dose scans to simulate 25\% of the full dose. This dataset serves as a gold standard for evaluating robustness against realistic, spatially correlated CT noise textures. For our experiments, we utilize the 1.0 mm slice reconstructions.

\paragraph{LIDC-IDRI (CT) \cite{Armato_2011_LIDC}:} To evaluate robustness under \textit{ultra-low-dose} conditions with severe streak artifacts, we specifically curated a subset of thoracic CT scans acquired with GE Medical Systems scanners. The selected data consists of high-resolution and standard scans with slice thicknesses of exactly 1.25 mm and 2.5 mm. We simulate ULD projections via the Radon Transform with a photon count of 12,500, followed by Filtered Back Projection (FBP). This generates realistic streak artifacts and non-stationary noise distributions that pose significant challenges to conventional SSL methods. Unless otherwise noted, aggregate LIDC-IDRI results pool the 1.25-mm and 2.5-mm test subsets; spacing-specific results are reported separately.
    
\paragraph{IXI-T1 (MRI) \cite{Biomedical_2018_IXI}:}
For a controlled cross-modality evaluation, we use brain scans from the IXI dataset with a slice thickness of 1.2\,mm. We add synthetic Rician noise at three intensity levels ($\sigma\in\{5\%,7\%,9\%\}$) to the clean volumetric data.

\subsubsection{Evaluation Metrics.}
We employ five quantitative metrics to assess different aspects of image quality. Peak Signal-to-Noise Ratio (PSNR) and Structural Similarity (SSIM) \cite{Wang_2004_SSIM} are used to measure general signal fidelity. To evaluate perceptual quality and low-level feature preservation, we utilize Feature Similarity (FSIM) \cite{Zhang_2011_FSIM}. Furthermore, given the importance of preserving fine anatomical details in medical imaging, we employ High-Frequency Error Norm (HFEN) \cite{Ravishankar_2011_HFEN} and Gradient Magnitude Similarity Deviation (GMSD) \cite{Xue_2014_GMSD} to specifically measure the restoration quality of edges and textures (lower values indicate better performance for HFEN and GMSD).

\subsubsection{Implementation Details.}
Our method is implemented in PyTorch and evaluated on a single CPU/GPU workstation (AMD Ryzen 9 9950X, NVIDIA RTX 5090). 
Baselines with public implementations follow released/recommended settings; NS-N2N \cite{Zhou_2024_NSN2N} is reimplemented from the paper using the same NAFNet \cite{Chen_2022_NAFNet} backbone as NGPS.
NGPS uses fixed default hyperparameters across datasets (patch size $p{=}7$, window size $W{=}15$, Top-$K$ $K{=}4$, masking threshold $\tau{=}0.05$), and we train with AdamW ($2{\times}10^{-4}$ learning rate, $10^{-5}$ weight decay) for 10 epochs with batch size 4, using $\lambda{=}0.5$.
Additional NGPS architecture and training details are provided in Supplementary Sec.~\ref{sec:supp_implementation}.

\begin{table}[t]
\centering
\caption{Quantitative evaluation on the AAPM-Mayo \cite{McCollough_2017_AAPM} (Quarter-Dose) and LIDC-IDRI \cite{Armato_2011_LIDC} (Ultra-Low-Dose) datasets. The best results are highlighted in \textbf{bold}, and the second-best are \underline{underlined}. $\uparrow$ indicates higher is better, while $\downarrow$ indicates lower is better.}
\label{tab:combined_ct_results}
\resizebox{\textwidth}{!}{%
\begin{tabular}{l|ccccc|ccccc}
\toprule
\multirow{2}{*}{\textbf{Method}} & \multicolumn{5}{c|}{\textbf{AAPM-Mayo (Quarter-Dose CT)}} & \multicolumn{5}{c}{\textbf{LIDC-IDRI (Simulated ULD CT)}} \\
\cmidrule(lr){2-6} \cmidrule(l){7-11}
 & \textbf{PSNR} $\uparrow$ & \textbf{SSIM} $\uparrow$ & \textbf{FSIM} $\uparrow$ & \textbf{HFEN} $\downarrow$ & \textbf{GMSD} $\downarrow$ & \textbf{PSNR} $\uparrow$ & \textbf{SSIM} $\uparrow$ & \textbf{FSIM} $\uparrow$ & \textbf{HFEN} $\downarrow$ & \textbf{GMSD} $\downarrow$ \\ 
\midrule
Baseline (Noisy) & 30.30 & 0.7222 & 0.8772 & 0.3363 & 0.0874 & 22.14 & 0.4213 & 0.5920 & 0.8429 & 0.1959 \\
\midrule
BM3D \cite{Dabov_2007_BM3D} & 34.67 & 0.7764 & 0.8547 & 0.3258 & 0.0900 & 25.07 & 0.5825 & 0.7768 & 0.6010 & 0.1449 \\
\midrule
DIP \cite{Ulyanov_2018_DIP} & 35.36 & 0.7895 & 0.8804 & 0.2766 & 0.0426 & 26.77 & 0.6235 & 0.8074 & 0.5910 & 0.1337 \\
NAC \cite{Xu_2020_NAC} & 34.61 & 0.8171 & 0.9062 & 0.2344 & 0.0664 & 24.83 & 0.5690 & 0.7500 & 0.6096 & 0.1439 \\
ZS-N2N \cite{Mansour_2023_ZSN2N} & 33.68 & 0.8129 & 0.9107 & 0.2911 & 0.0631 & 26.63 & 0.5979 & 0.8080 & 0.5545 & 0.1259 \\
Pixel2Pixel \cite{Ma_2025_Pixel2Pixel} & 33.71 & 0.8357 & 0.9117 & 0.3183 & 0.0739 & 25.58 & 0.5078 & 0.7277 & 0.6196 & 0.1237 \\
\midrule
Noise2Void \cite{Krull_2019_N2V} & 32.57 & 0.7786 & 0.8977 & 0.3286 & 0.0739 & 26.10 & 0.5995 & 0.8295 & 0.4948 & 0.1078 \\
NB2NB \cite{Huang_2021_NB2NB} & 33.21 & 0.7922 & 0.9037 & 0.3187 & 0.0715 & 27.23 & 0.6063 & 0.8321 & 0.4687 & 0.1003 \\
Filter2Noise \cite{Sun_2025_Filter2Noise} & 35.23 & 0.8334 & 0.9265 & 0.2895 & 0.0625 & 28.81 & 0.7181 & 0.8716 & 0.5188 & 0.1036 \\
\midrule
Deformed2Self \cite{Xu_2021_Deformed2Self} & 35.85 & \underline{0.8662} & 0.9144 & \underline{0.2299} & \underline{0.0377} & 28.94 & 0.6703 & 0.8170 & 0.5134 & 0.0808 \\ 
Noise2Sim \cite{Niu_2023_Noise2Sim} & 35.49 & 0.8639 & \underline{0.9420} & 0.2406 & 0.0390 & 28.81 & 0.7817 & 0.8749 & 0.5003 & 0.1040 \\
NS-N2N \cite{Zhou_2024_NSN2N} & \underline{35.91} & 0.8584 & 0.9235 & 0.2325 & 0.0396 & \underline{30.62} & \underline{0.8080} & \underline{0.8944} & \underline{0.4406} & \textbf{0.0777} \\
\rowcolor{gray!10}
\textbf{Ours} & \textbf{36.68} & \textbf{0.8986} & \textbf{0.9470} & \textbf{0.2056} & \textbf{0.0362} & \textbf{31.03} & \textbf{0.8102} & \textbf{0.9168} & \textbf{0.4161} & \underline{0.0788} \\ 
\bottomrule
\end{tabular}%
}
\end{table}

\subsection{Quantitative Comparison}

\subsubsection{Results on Low-Dose and Ultra-Low-Dose CT Datasets.}

Table~\ref{tab:combined_ct_results} compares quarter-dose AAPM-Mayo \cite{McCollough_2017_AAPM} and simulated-ULD LIDC-IDRI \cite{Armato_2011_LIDC}.
NGPS leads all five AAPM metrics.
On LIDC-IDRI, it improves PSNR, FSIM, and HFEN over NS-N2N \cite{Zhou_2024_NSN2N} by 0.41\,dB, 0.0224, and 0.0245, respectively, whereas the SSIM margin is small (0.0022) and NS-N2N has a slightly lower GMSD (0.0777 vs.\ 0.0788).
Volume-level paired 95\% confidence intervals exclude zero for PSNR and FSIM, but include zero for SSIM, HFEN, and GMSD (Supplementary Table~\ref{tab:supp_lidc_paired_ci}).
We therefore interpret the LIDC-IDRI result as improved fidelity and boundary-sensitive restoration rather than uniform dominance across all metrics.
In the evaluated ULD and thicker-slice settings, Deformed2Self \cite{Xu_2021_Deformed2Self} shows larger degradation, while masking-based methods omit discrepant regions from their reconstruction losses.
NGPS instead retrieves local supervision in these regions without dense warping.

\begin{figure}[t]
\centering
\includegraphics[width=0.9\textwidth]{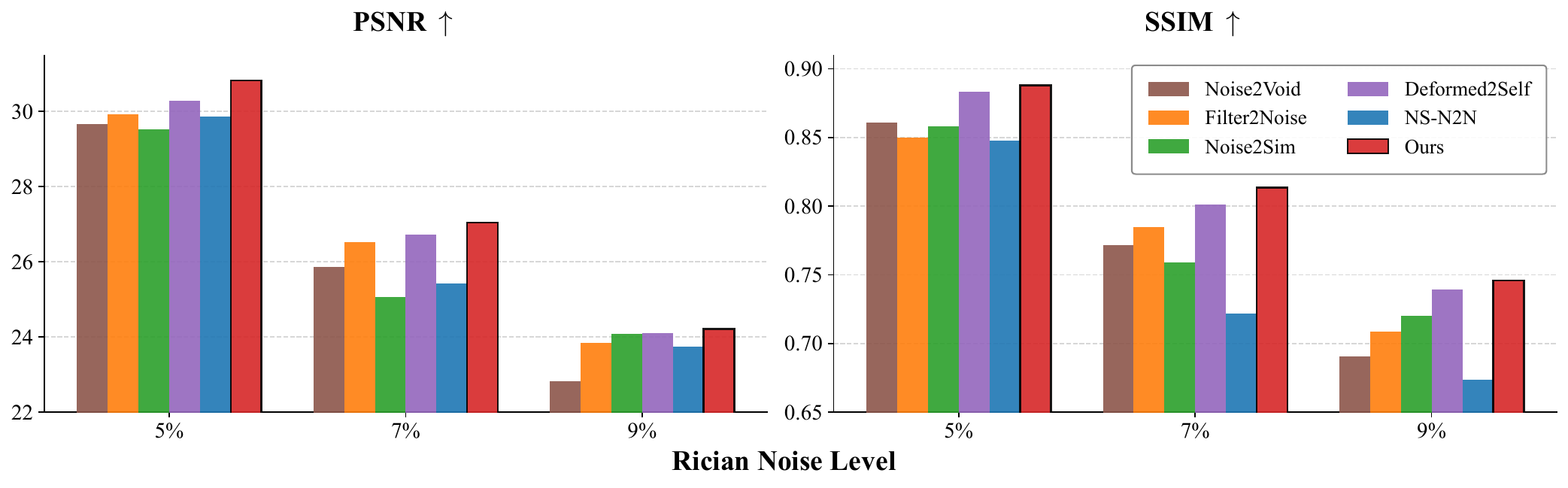}
\caption{\textbf{Quantitative comparison on IXI dataset \cite{Biomedical_2018_IXI} (Simulated Rician noise).} We add different levels of Rician noise (5\%, 7\%, and 9\%) for corruption.}
\label{fig:ixi_results}
\end{figure}

\subsubsection{Results on the IXI MRI Dataset with Rician Noise.}
We evaluate robustness on the IXI dataset \cite{Biomedical_2018_IXI} under increasing synthetic Rician noise (5\%, 7\%, and 9\%), summarizing performance trends in Fig.~\ref{fig:ixi_results} (full values in Supplementary Table~\ref{tab:supp_ixi_results}).
All methods degrade as the corruption level increases, but NGPS maintains the highest PSNR and SSIM across the tested levels. Among the baselines, Deformed2Self \cite{Xu_2021_Deformed2Self} ranks second overall, which is consistent with the usefulness of inter-slice context in this setting. The relative gaps among Deformed2Self, Noise2Sim \cite{Niu_2023_Noise2Sim}, NS-N2N \cite{Zhou_2024_NSN2N}, and the single-image baselines vary across corruption levels. Nevertheless, NGPS remains highest in both metrics throughout the tested range, demonstrating consistent performance under the evaluated synthetic Rician corruption.

\begin{figure}[t]
\centering
\begin{subfigure}[h]{\textwidth}
  \centering
  \includegraphics[width=\textwidth]{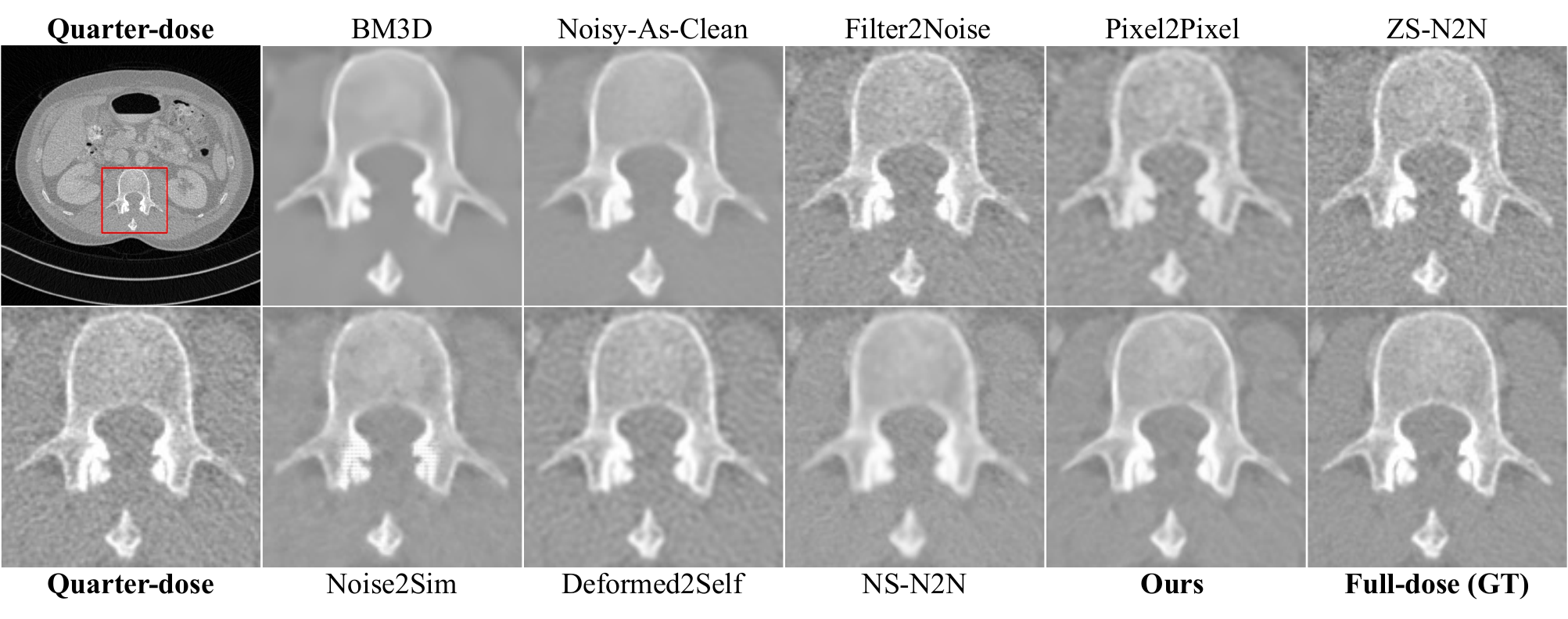}
  \caption{AAPM-Mayo \cite{McCollough_2017_AAPM} (Quarter-Dose CT). Pronounced quantum noise and over-smoothing artifacts.}
  \label{fig:qual_aapm}
\end{subfigure}

\begin{subfigure}[h]{\textwidth}
  \centering
  \includegraphics[width=\textwidth]{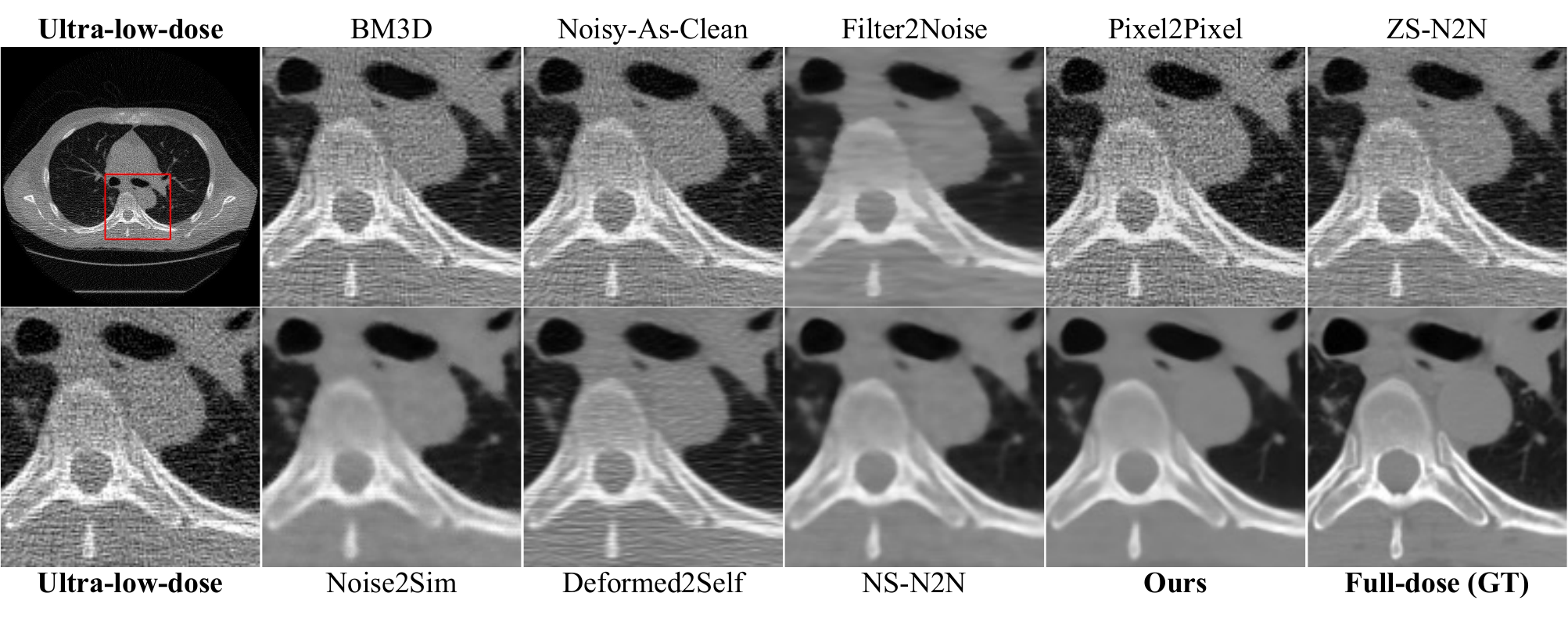}
  \caption{LIDC-IDRI \cite{Armato_2011_LIDC} (Simulated ULD CT, $P{=}12.5K$). Severe, spatially correlated streak artifacts.}
  \label{fig:qual_lidc}
\end{subfigure}
\caption{\textbf{Qualitative comparisons on quarter-dose and ultra-low-dose CT.} The top-left shows the noisy input with a red ROI, while the rest display magnified ROIs.}
\label{fig:ct_qualitative}
\end{figure}

\subsection{Qualitative Comparison}
Fig.~\ref{fig:ct_qualitative} compares quarter-dose AAPM-Mayo \cite{McCollough_2017_AAPM} and ULD LIDC-IDRI \cite{Armato_2011_LIDC} CTs.
The baselines exhibit clear visual trade-offs in the highlighted ROIs: single-image methods such as BM3D \cite{Dabov_2007_BM3D}, Noisy-As-Clean \cite{Xu_2020_NAC}, Filter2Noise \cite{Sun_2025_Filter2Noise}, Pixel2Pixel \cite{Ma_2025_Pixel2Pixel}, and ZS-N2N \cite{Mansour_2023_ZSN2N} either retain more residual noise or produce softer anatomical details.
The warping-based baseline Deformed2Self \cite{Xu_2021_Deformed2Self} shows local artifacts in some regions, consistent with the difficulty of estimating correspondence under strong degradation.
Masking-based SSL methods such as Noise2Sim \cite{Niu_2023_Noise2Sim} and NS-N2N \cite{Zhou_2024_NSN2N} produce smoother boundaries in the displayed ROIs, which is consistent with omitting discrepant pixels from the reconstruction loss.
In these examples, NGPS reduces residual corruption while retaining sharper anatomical margins through local raw-target retrieval.

\begin{figure}[t]
\centering
\includegraphics[width=\textwidth]{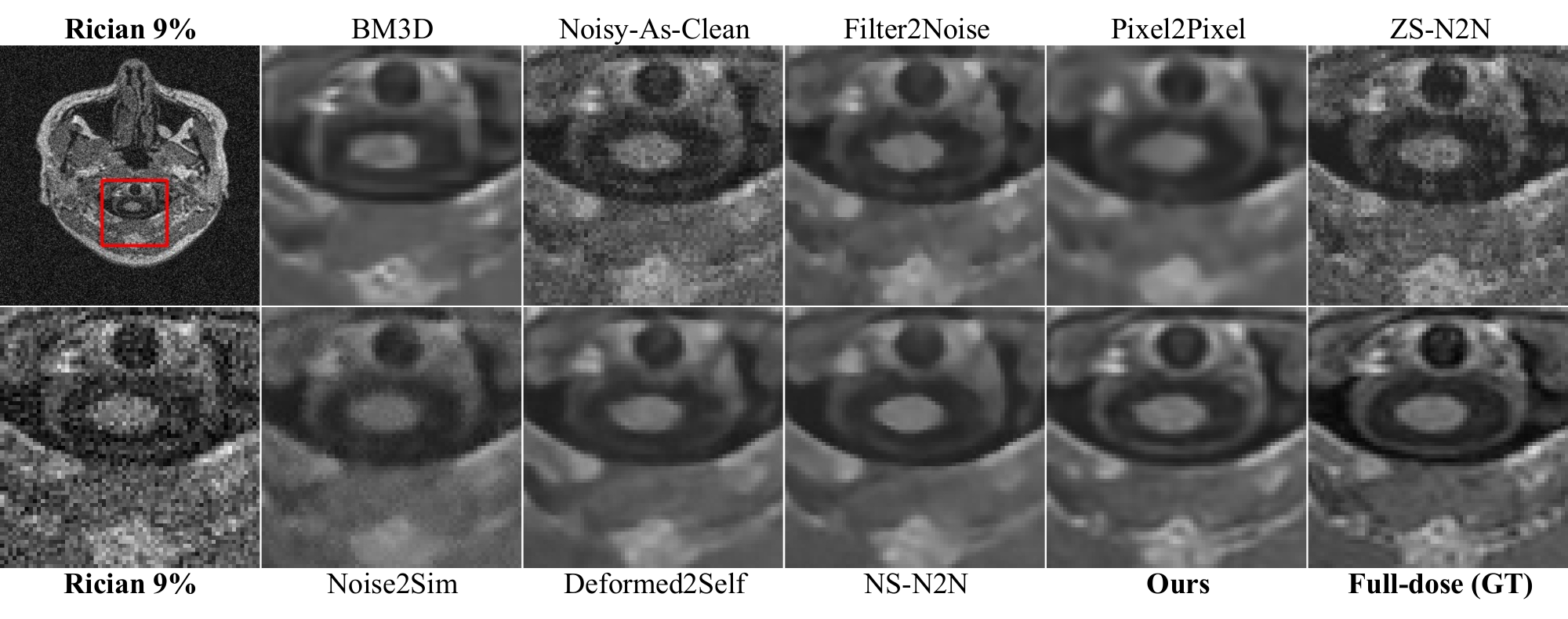}
\caption{\textbf{Qualitative results on IXI \cite{Biomedical_2018_IXI} under 9\% synthetic Rician noise.}
In the highlighted region, several baselines show more residual corruption or softer cortical boundaries.
The warping- and masking-based baselines exhibit local structure loss in this example, whereas NGPS retains sharper anatomical details.}
\label{fig:ixi_qualitative}
\end{figure}

Under 9\% synthetic Rician noise on IXI MRI \cite{Biomedical_2018_IXI} (Fig.~\ref{fig:ixi_qualitative}), similar tendencies are visible.
Single-image methods retain more residual corruption or oversmoothing, the warping-based baseline shows local degradation under the tested corruption, and masking-based outputs show softer cortical boundaries in the highlighted area. In the highlighted example, NGPS retains sharper local structures while reducing residual corruption, consistent with the quantitative trends.

\begin{table}[t]
\centering
\caption{Sensitivity to slice thickness on LIDC-IDRI \cite{Armato_2011_LIDC}. $\Delta$ PSNR reports the change from 1.25\,mm to 2.5\,mm; the best results are highlighted in \textbf{bold}.}
\label{tab:thickness_ablation}
\resizebox{0.85\textwidth}{!}{%
\begin{tabular}{l|cc|cc|c}
\toprule
\multirow{2}{*}{\textbf{Method}} & \multicolumn{2}{c|}{\textbf{1.25 mm (Thin Slice)}} & \multicolumn{2}{c|}{\textbf{2.5 mm (Thick Slice)}} & \multirow{2}{*}{\textbf{$\Delta$ PSNR $\downarrow$}} \\
\cmidrule{2-5}
& \textbf{PSNR} $\uparrow$ & \textbf{SSIM} $\uparrow$ & \textbf{PSNR} $\uparrow$ & \textbf{SSIM} $\uparrow$ & \\
\midrule
Noise2Sim \cite{Niu_2023_Noise2Sim} & 30.53 & 0.7963 & 27.70 & 0.7738 & -2.83 dB \\
Deformed2Self \cite{Xu_2021_Deformed2Self} & 30.52 & 0.7553 & 28.64 & 0.6550 & -1.88 dB \\
NS-N2N \cite{Zhou_2024_NSN2N} & 30.83 & 0.8092 & 30.20 & 0.7877 & -0.63 dB \\
\rowcolor{gray!10}
\textbf{Ours (NGPS)} & \textbf{31.08} & \textbf{0.8103} & \textbf{30.92} & \textbf{0.7994} & \textbf{-0.16 dB} \\
\bottomrule
\end{tabular}%
}
\end{table}

\subsection{Discussion}

\subsubsection{Robustness to Slice Thickness and Masking Threshold.}

Table~\ref{tab:thickness_ablation} isolates spacing changes using independently trained 1.25\,mm and 2.5\,mm models.
NGPS decreases by 0.16\,dB, compared with 0.63\,dB for NS-N2N \cite{Zhou_2024_NSN2N}, 1.88\,dB for Deformed2Self \cite{Xu_2021_Deformed2Self}, and 2.83\,dB for Noise2Sim \cite{Niu_2023_Noise2Sim}.
The threshold sweep in Fig.~\ref{fig:threshold_ablation} further shows that NGPS is comparatively stable across $\tau$, whereas NS-N2N degrades under both strict and loose masking.
These results support robustness within the evaluated spacing and threshold ranges; controlled larger-gap failures and confidence-based rejection are analyzed in Supplementary Sec.~\ref{sec:supp_gap_stress}.

\subsubsection{Multi-geometry Search-window Selection.}

We estimate match-offset CDFs using a $41{\times}41$ reference search across AAPM \cite{McCollough_2017_AAPM} 1.0\,mm, IXI \cite{Biomedical_2018_IXI} 1.2\,mm, and LIDC-IDRI \cite{Armato_2011_LIDC} 1.25/2.5\,mm, with $p{=}7$ and $K{=}4$.
A $15{\times}15$ window covers over 80\% of the selected offsets for every geometry and 98\% on LIDC-IDRI 1.25\,mm (Fig.~\ref{fig:window_ablation}).
Because larger windows add quadratic candidate cost with limited coverage gain, we use $W{=}15$ as a common default.
Supplementary Table~\ref{tab:supp_window_spacing} shows that $W{=}15$ is within 0.02\,dB of the best 1.25\,mm result and performs best at 2.5\,mm; we therefore use it as a practical common default rather than a universal optimum.

\subsubsection{Effect of Top-$K$ Ensemble Size.}
\begin{figure}[!t]
\centering
\includegraphics[width=0.9\textwidth]{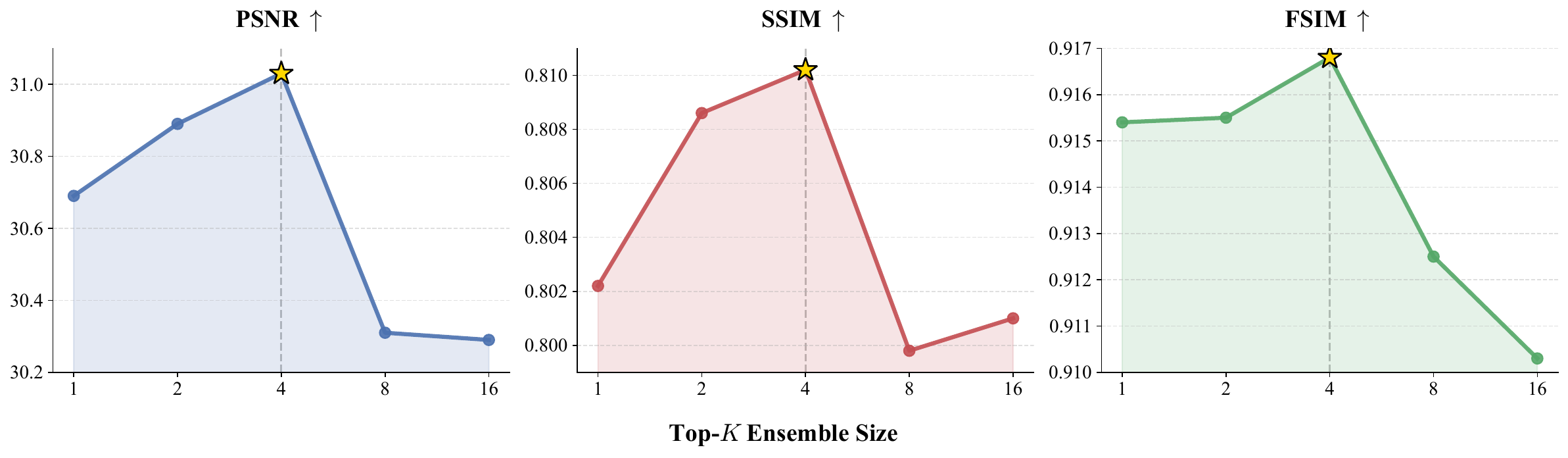}
\caption{\textbf{Ablation study} on the Top-$K$ ensemble size in the NGPS module. }
\label{fig:ablation_k}
\end{figure}

In NGPS, $K$ controls the number of retrieved candidate patches used to form the Top-$K$ ensemble target.
Fig.~\ref{fig:ablation_k} shows a clear trade-off on the LIDC dataset \cite{Armato_2011_LIDC}: performance improves from small ensembles to a moderate $K$, then declines when $K$ becomes too large. The best results are achieved at $K=4$ (PSNR 31.03\,dB, SSIM 0.8102). With $K=1$ or $2$, the ensemble is too small to sufficiently reduce residual noise. With larger $K$ ($8$ or $16$), lower-quality or misaligned matches are increasingly included, which blurs fine structures. Thus, $K=4$ provides the best tested noise-detail balance.

\subsubsection{Computational Efficiency.}
Table~\ref{tab:time_overhead} compares target-preparation cost for a 100-slice volume. NGPS requires approximately 0.72\,s, about $19\times$ faster than NS-N2N \cite{Zhou_2024_NSN2N} and $7.5\times$ faster than Pixel2Pixel \cite{Ma_2025_Pixel2Pixel}; Noise2Sim \cite{Niu_2023_Noise2Sim} is faster but attains lower restoration metrics in Table~\ref{tab:combined_ct_results}. The efficiency of NGPS comes from restricting vectorized patch search to masked pixels and using the lightweight bilateral-plus-median guide.
\FloatBarrier
\setcounter{topnumber}{3}
\setcounter{totalnumber}{4}
\begin{figure}[!t]
\centering
\begin{subfigure}[t]{0.45\textwidth}
  \centering
  \includegraphics[width=\textwidth]{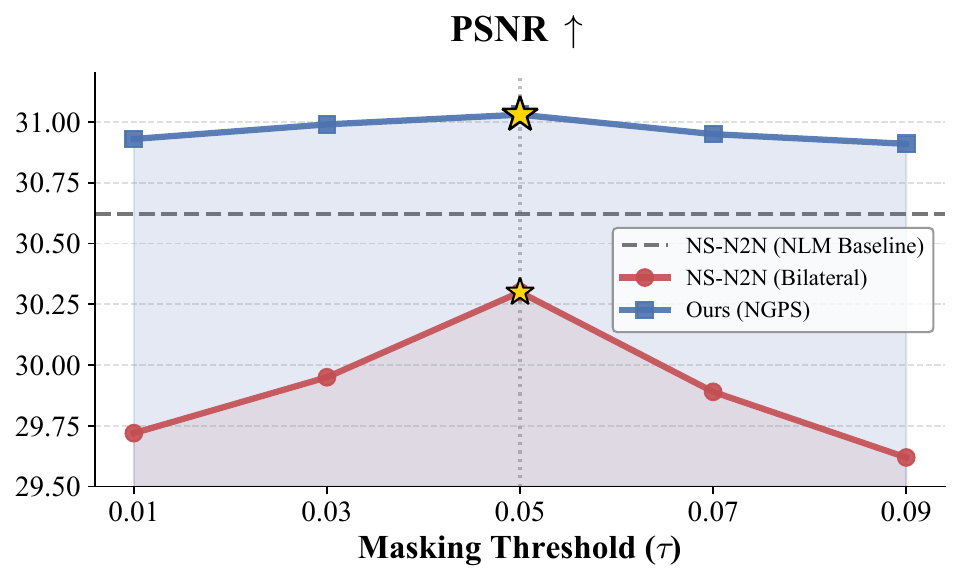}
  \caption{Ablation of Masking threshold $\tau$.}
  \label{fig:threshold_ablation}
\end{subfigure}\hfill
\begin{subfigure}[t]{0.55\textwidth}
  \centering
  \includegraphics[width=\textwidth]{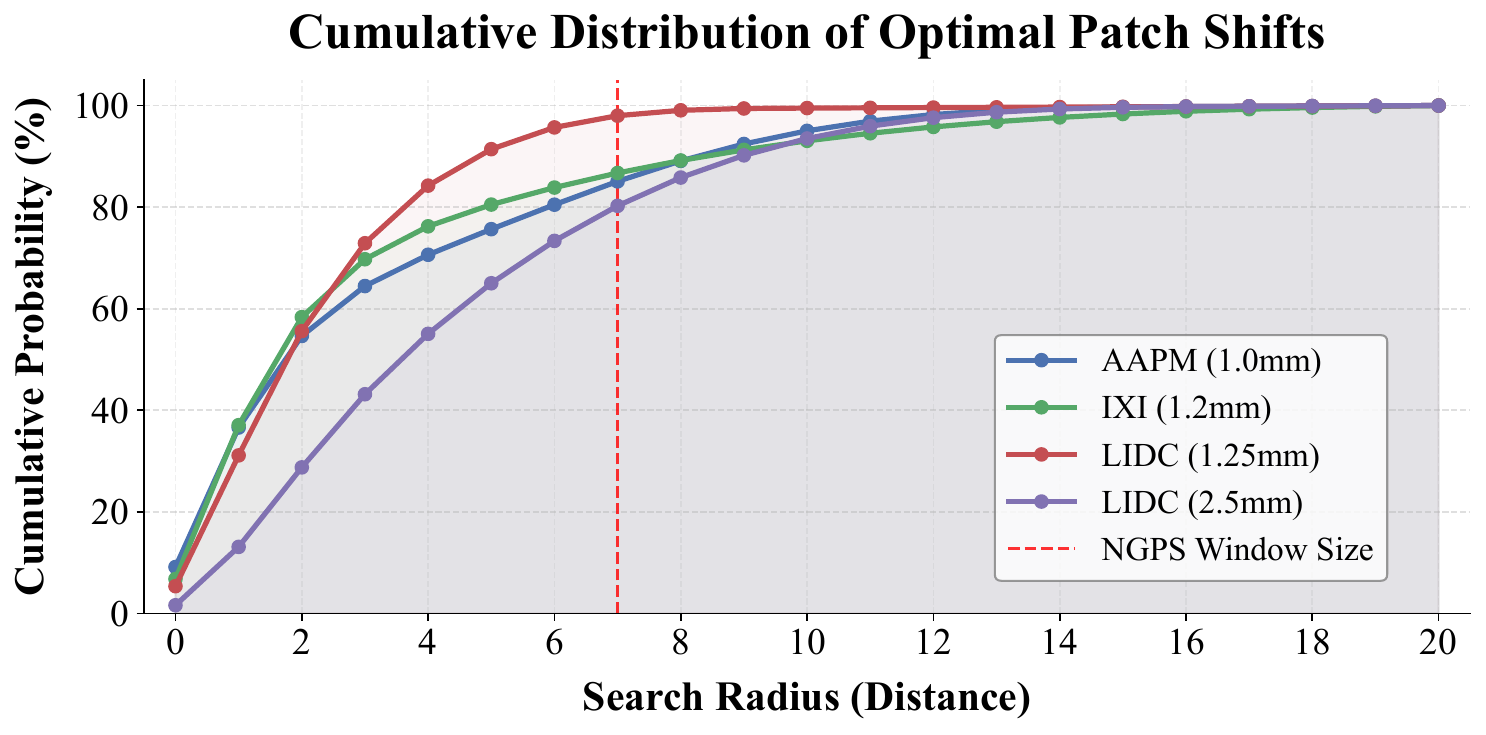}
  \caption{CDF of inter-slice displacement (window selection).}
  \label{fig:window_ablation}
\end{subfigure}
\caption{\textbf{Ablations for mask sensitivity and search-window design.} Left: PSNR trends under varying $\tau$. Right: displacement CDF used to set the NGPS search radius.}
\label{fig:mask_window_ablations}
\end{figure}

\begin{table}[!t]
\centering
\caption{Comparison of preprocessing time overhead for misalignment handling. Time is measured for a standard medical volume consisting of 100 slices ($512 \times 512$ resolution).}
\label{tab:time_overhead}
\resizebox{\textwidth}{!}{%
\begin{tabular}{l|l|c|c}
\toprule
\textbf{Method} & \textbf{Target Preparation Strategy} & \textbf{Compute Device} & \textbf{Time (per 100 slices)} $\downarrow$ \\
\midrule
Pixel2Pixel \cite{Ma_2025_Pixel2Pixel} & Patch-based Pixel Bank Creation & GPU & $\sim$5.38 s \\
Noise2Sim \cite{Niu_2023_Noise2Sim} & Mean Filter Masking & GPU & \textbf{$\sim$0.005 s} \\
NS-N2N \cite{Zhou_2024_NSN2N} & NLM Denoising + Median Masking & CPU + GPU & $\sim$13.36 s \\
\rowcolor{gray!10}
\textbf{Ours (NGPS)} & LPF Masking + Vectorized Patch Search & CPU + GPU & \underline{$\sim$0.72 s} \\
\bottomrule
\end{tabular}%
}
\end{table}
\section{Conclusion}
We presented NGPS, a lightweight framework for constructing neighboring-slice supervision under inter-slice misalignment. NGPS forms noise-attenuated guides and direction-specific discrepancy masks, performs local Top-$K$ patch search at flagged locations, and retrieves the corresponding raw adjacent-slice values as training targets. Its hybrid objective combines same-coordinate neighboring supervision at unflagged locations with NGPS-retrieved targets at flagged locations, together with regional consistency, without dense warping or a learnable alignment module. Across a realistic quarter-dose CT benchmark, simulated ultra-low-dose CT, and synthetic-Rician MRI, NGPS consistently improves fidelity and structure-sensitive restoration: it leads all AAPM \cite{McCollough_2017_AAPM} metrics, improves PSNR, FSIM, and HFEN over NS-N2N \cite{Zhou_2024_NSN2N} on simulated ULD CT, and achieves the highest PSNR and SSIM across the evaluated MRI noise levels. These results support local supervision retrieval as an effective and practical strategy for structure-preserving volumetric self-supervised denoising.

\paragraph{Limitations and Future Work.}
NGPS assumes that a locally corresponding structure with sufficiently consistent appearance exists within the search window. This assumption can fail under large through-plane gaps, abrupt anatomical changes, staining variation, signal dropout, or other non-geometric inter-slice appearance changes. The optional match-cost gate mitigates degradation in the tested large-gap setting, but does not remove this assumption. The Noise2Noise \cite{Lehtinen_2018_N2N} formulation does not require pixel-wise i.i.d.\ noise within each slice, but assumes zero-mean noise and negligible cross-slice correlation between input- and target-slice noise; cross-slice correlated artifacts remain outside the current validation scope. The fixed $p$, $W$, $K$, and $\tau$ values are practical defaults rather than universal optima. Finally, the IXI \cite{Biomedical_2018_IXI} experiments provide a controlled synthetic-noise evaluation; future work should prioritize acquisition-realistic low-field MRI, noisy 3D microscopy, and downstream clinical validation.

\section*{Acknowledgements}
This work was supported by the SNUAILAB, the Institute of Information $\&$ Communications Technology Planning $\&$ Evaluation (IITP) grant funded by the Korea government (MSIT) [RS-2021-II211341, Artificial Intelligence Graduate School Program (Chung-Ang University) and RS-2022-II220124, Development of Artificial Intelligence Technology for Self-Improving Competency-Aware Learning Capabilities]. Also, this research was supported by the "Regional Innovation System \& Education
(RISE)" through the Seoul RISE Center, funded by the Ministry of Education (MOE)
and the Seoul Metropolitan Government. (2026-RISE-01-024-04)

\bibliographystyle{splncs04}
\bibliography{main}

\clearpage

\setcounter{section}{0}
\setcounter{subsection}{0}
\setcounter{subsubsection}{0}
\setcounter{equation}{0}
\setcounter{figure}{0}
\setcounter{table}{0}
\setcounter{algorithm}{0}

\renewcommand{\thesection}{S\arabic{section}}
\renewcommand{\thesubsection}{\thesection.\arabic{subsection}}
\renewcommand{\thesubsubsection}{\thesubsection.\arabic{subsubsection}}
\renewcommand{\theequation}{S\arabic{equation}}
\renewcommand{\thefigure}{S\arabic{figure}}
\renewcommand{\thetable}{S\arabic{table}}
\renewcommand{\thealgorithm}{S\arabic{algorithm}}

\renewcommand{\theHsection}{supp.\arabic{section}}
\renewcommand{\theHsubsection}{supp.\arabic{section}.\arabic{subsection}}
\renewcommand{\theHsubsubsection}{supp.\arabic{section}.\arabic{subsection}.\arabic{subsubsection}}
\renewcommand{\theHequation}{supp.\arabic{equation}}
\renewcommand{\theHfigure}{supp.\arabic{figure}}
\renewcommand{\theHtable}{supp.\arabic{table}}
\providecommand{\theHalgorithm}{}
\renewcommand{\theHalgorithm}{supp.\arabic{algorithm}}

\markboth{Supplementary Material}{Supplementary Material}

\section{Detailed Implementation Details}
\label{sec:supp_implementation}

\begin{algorithm}[htbp]
\caption{Neighbor-Guided Patch Sampling (NGPS)}
\label{alg:ngps}
\textbf{Input:} Noisy volume $y \in \mathbb{R}^{H \times W \times D}$, Denoiser $f_\theta$\\
\textbf{Parameters:} Mask threshold $\tau$, search window $\Omega$, ensemble size $K$, RC weight $\lambda$, learning rate $\eta$

\begin{algorithmic}[1]
\STATE \textbf{\% Phase 1: Fast Guide Generation \& Masking}
\STATE $\tilde{y} \leftarrow \mathrm{MF}(\mathrm{BF}(y))$ \hfill \COMMENT{Edge-preserving filtering}%
\FOR{each slice $z$ and adjacent $z' \in \{z-1, z+1\}$}
    \STATE $\mathcal{M}_{z \to z'} \leftarrow \mathbf{1}\big(|\tilde{y}_z - \tilde{y}_{z'}| > \tau\big)$ \hfill \COMMENT{1: Misaligned, 0: Static}
\ENDFOR

\vspace{0.1cm}
\STATE \textbf{\% Phase 2: Decoupled Matching \& Retrieval}
\FOR{each slice $z$ and adjacent $z'$}
    \FOR{each misaligned pixel $p$ where $\mathcal{M}_{z \to z'}(p) = 1$}
        \STATE \COMMENT{Match on guide $\tilde{y}$}
        \STATE $\{q^{(k)}\}_{k=1}^K \leftarrow \arg\text{Top-}K_{q \in \Omega_p} \big\| \mathcal{P}(\tilde{y}_z, p) - \mathcal{P}(\tilde{y}_{z'}, q) \big\|_2^2$ 
        \STATE \COMMENT{Retrieve from raw $y$}
        \STATE $t_{z \to z'}(p) \leftarrow \frac{1}{K}\sum_{k=1}^{K} y_{z'}\big(q^{(k)}\big)$ 
    \ENDFOR
\ENDFOR

\vspace{0.1cm}
\STATE \textbf{\% Phase 3: Hybrid Objective Training}
\FOR{$e=1,\ldots,10$}
    \FOR{each mini-batch of slice triplets
    $\{y_z,y_{z-1},y_{z+1}\}$}
        \STATE Predict
        $\hat{x}_s \leftarrow f_\theta(y_s)$,
        $s\in\{z,z-1,z+1\}$

        \STATE Calculate
        $\mathcal{L}_{N2N}$,
        $\mathcal{L}_{NGPS}$, and
        $\mathcal{L}_{RC}$

        \STATE Update
        $\theta \leftarrow
        \theta-\eta\nabla_\theta
        (\mathcal{L}_{N2N}
        +\mathcal{L}_{NGPS}
        +\lambda\mathcal{L}_{RC})$
    \ENDFOR
\ENDFOR
\RETURN Optimized weights $\theta^*$
\end{algorithmic}
\end{algorithm}

To provide a clear, step-by-step mathematical overview of our proposed framework, Algorithm~\ref{alg:ngps} summarizes the entire training pipeline. The algorithm is logically divided into three phases: edge-preserving guide generation, direction-aware target retrieval, and hybrid objective training. The specific hyperparameters utilized within this algorithm are detailed in the following subsections.

\subsection{Network Architecture}
Because our contribution concerns supervision construction rather than network
architecture, we use NAFNet \cite{Chen_2022_NAFNet} as the denoising backbone.
The base width is 32, with encoder blocks $[2,2,4,8]$, eight middle blocks,
and decoder blocks $[2,2,2,2]$.

\subsection{Optimization and Training Configuration}
The framework was implemented using PyTorch and trained on a workstation equipped with an AMD Ryzen 9 9950X CPU and a single NVIDIA RTX 5090 GPU. The model was optimized using the AdamW optimizer with an initial learning rate of $2 \times 10^{-4}$ and a weight decay of $10^{-5}$. During training, spatial dimensions were randomly cropped to $256 \times 256$ to serve as network inputs. We trained the network for 10 epochs with a batch size of 4 across all datasets. The regional-consistency weight was set to $\lambda_{rc}=0.5$ in all experiments.

\subsection{NGPS Module and Guide Generation}
We use $p=7$, $W=15$, $K=4$, and $\tau=0.05$ as common defaults across the evaluated datasets. Geometry-specific window sensitivity is reported in Table~\ref{tab:supp_window_spacing}. 

To compute the structural discrepancies and flag misaligned pixels, we generate a noise-attenuated guide volume using a sequential combination of a Bilateral filter and a Median filter. For the Bilateral filter, we set the spatial window size to $d=5$, color sigma to $\sigma_{color}=35$, and space sigma to $\sigma_{space}=50$. Specifically, to ensure a fair and direct comparison with the passive masking baseline, the kernel size of the Median filter was set to $5 \times 5$. The $5\times5$ median kernel follows the NS-N2N configuration \cite{Zhou_2024_NSN2N}; the preceding bilateral stage is the lightweight
guide choice evaluated for NGPS. Following this lightweight filtering, the discrepancy masking threshold was fixed at $\tau=0.05$.

\section{Additional Quantitative Results}
\label{sec:supp_quantitative}

\begin{table}[t]
\centering
\caption{Detailed quantitative evaluation on the IXI dataset \cite{Biomedical_2018_IXI}. The best results are highlighted in \textbf{bold}, and the second-best are \underline{underlined}. $\uparrow$ indicates higher is better.}
\label{tab:supp_ixi_results}
\resizebox{\textwidth}{!}{%
\begin{tabular}{l|cc|cc|cc}
\toprule
\multirow{2}{*}{\textbf{Method}} & \multicolumn{2}{c|}{\textbf{5\% Noise}} & \multicolumn{2}{c|}{\textbf{7\% Noise}} & \multicolumn{2}{c}{\textbf{9\% Noise}} \\
\cmidrule(lr){2-3} \cmidrule(lr){4-5} \cmidrule(l){6-7}
 & \textbf{PSNR} $\uparrow$ & \textbf{SSIM} $\uparrow$ & \textbf{PSNR} $\uparrow$ & \textbf{SSIM} $\uparrow$ & \textbf{PSNR} $\uparrow$ & \textbf{SSIM} $\uparrow$ \\ 
\midrule
Baseline (Rician) & 25.93 & 0.5276 & 22.71 & 0.4070 & 20.29 & 0.3255 \\
\midrule
BM3D \cite{Dabov_2007_BM3D} & 28.77 & 0.8503 & 25.53 & 0.7789 & 23.05 & 0.7154 \\
\midrule
DIP \cite{Ulyanov_2018_DIP} & 29.51 & 0.8118 & 25.25 & 0.6390 & 22.15 & 0.4967 \\
NAC \cite{Xu_2020_NAC} & 29.39 & 0.8107 & 25.65 & 0.6568 & 23.13 & 0.5889 \\
ZS-N2N \cite{Mansour_2023_ZSN2N} & 29.20 & 0.8381 & 25.74 & 0.7464 & 23.21 & 0.6723 \\
Pixel2Pixel \cite{Ma_2025_Pixel2Pixel} & 29.18 & 0.8444 & 25.99 & 0.7634 & 23.48 & 0.6953 \\ 
\midrule
Noise2Void \cite{Krull_2019_N2V} & 29.66 & \underline{0.8609} & 25.86 & 0.7717 & 22.82 & 0.6903 \\
NB2NB \cite{Huang_2021_NB2NB} & 29.35 & 0.8400 & 26.42 & 0.7663 & 23.82 & 0.6939 \\
Filter2Noise \cite{Sun_2025_Filter2Noise} & \underline{29.93} & 0.8502 & 26.53 & 0.7850 & 23.84 & 0.7086 \\ 
\midrule
Deformed2Self \cite{Xu_2021_Deformed2Self} & 30.28 & 0.8831 & \underline{26.73} & \underline{0.8011} & \underline{24.11} & \underline{0.7392} \\
Noise2Sim \cite{Niu_2023_Noise2Sim} & 29.52 & 0.8579 & 25.06 & 0.7589 & 24.08 & 0.7200 \\
NS-N2N \cite{Zhou_2024_NSN2N} & 29.86 & 0.8478 & 25.42 & 0.7215 & 23.74 & 0.6735 \\
\rowcolor{gray!10}
\textbf{Ours} & \textbf{30.83} & \textbf{0.8879} & \textbf{27.04} & \textbf{0.8135} & \textbf{24.21} & \textbf{0.7459} \\ 
\bottomrule
\end{tabular}%
}
\end{table}

\subsection{Detailed Results on the IXI Dataset}
In the main manuscript, we summarized the denoising performance on the IXI MRI dataset \cite{Biomedical_2018_IXI} under simulated Rician noise using bar-chart histograms to illustrate the robustness trends. For completeness and precise numerical comparison, Table~\ref{tab:supp_ixi_results} provides the quantitative measurements (PSNR and SSIM) for all evaluated methods across the noise intensity levels (5\%, 7\%, and 9\%). 

Across the evaluated 5\%, 7\%, and 9\% synthetic Rician settings, NGPS attains the highest PSNR and SSIM, with Deformed2Self \cite{Xu_2021_Deformed2Self} ranking second overall. These results characterize robustness to the evaluated synthetic corruption model.

\begin{table}[t]
\centering
\caption{Robustness analysis of different denoising methods on the LIDC-IDRI dataset \cite{Armato_2011_LIDC} across three random seeds. Results are reported as Mean $\pm$ Standard Deviation. BM3D is deterministic and thus has no variance. The best mean results are highlighted in \textbf{bold}, and the second-best are \underline{underlined}.}
\label{tab:seed_robustness}
\resizebox{\textwidth}{!}{%
\begin{tabular}{l|ccccc}
\toprule
\textbf{Method} & \textbf{PSNR} $\uparrow$ & \textbf{SSIM} $\uparrow$ & \textbf{FSIM} $\uparrow$ & \textbf{HFEN} $\downarrow$ & \textbf{GMSD} $\downarrow$ \\ 
\midrule
BM3D \cite{Dabov_2007_BM3D} & 25.07 & 0.5825 & 0.7768 & 0.6010 & 0.1449 \\
\midrule
DIP \cite{Ulyanov_2018_DIP} & 26.77 $\pm$ 0.0390 & 0.6235 $\pm$ 0.0016 & 0.8074 $\pm$ 0.0022 & 0.5910 $\pm$ 0.0013 & 0.1337 $\pm$ 0.0005 \\
NAC \cite{Xu_2020_NAC} & 24.83 $\pm$ 0.0605 & 0.5690 $\pm$ 0.0082 & 0.7500 $\pm$ 0.0018 & 0.6096 $\pm$ 0.0039 & 0.1439 $\pm$ 0.0010 \\
ZS-N2N \cite{Mansour_2023_ZSN2N} & 26.63 $\pm$ 0.0129 & 0.5979 $\pm$ 0.0006 & 0.8080 $\pm$ 0.0006 & 0.5545 $\pm$ 0.0009 & 0.1259 $\pm$ 0.0003 \\
Pixel2Pixel \cite{Ma_2025_Pixel2Pixel} & 25.58 $\pm$ 0.0213 & 0.5078 $\pm$ 0.0017 & 0.7277 $\pm$ 0.0003 & 0.6196 $\pm$ 0.0005 & 0.1237 $\pm$ 0.0001 \\
\midrule
Noise2Void \cite{Krull_2019_N2V} & 26.10 $\pm$ 0.2462 & 0.5995 $\pm$ 0.0767 & 0.8295 $\pm$ 0.0046 & 0.4948 $\pm$ 0.0089 & 0.1078 $\pm$ 0.0040 \\
NB2NB \cite{Huang_2021_NB2NB} & 27.23 $\pm$ 0.2045 & 0.6063 $\pm$ 0.0276 & 0.8321 $\pm$ 0.0039 & 0.4687 $\pm$ 0.0073 & 0.1003 $\pm$ 0.0030 \\
Filter2Noise \cite{Sun_2025_Filter2Noise} & 28.81 $\pm$ 0.1260 & 0.7181 $\pm$ 0.0045 & 0.8716 $\pm$ 0.0041 & 0.5188 $\pm$ 0.0208 & 0.1036 $\pm$ 0.0025 \\
\midrule
Deformed2Self \cite{Xu_2021_Deformed2Self} & 28.94 $\pm$ 0.0171 & 0.6703 $\pm$ 0.0015 & 0.8170 $\pm$ 0.0005 & 0.5134 $\pm$ 0.0003 & 0.0808 $\pm$ 0.0001 \\ 
Noise2Sim \cite{Niu_2023_Noise2Sim} & 28.81 $\pm$ 0.0952 & 0.7817 $\pm$ 0.0046 & 0.8749 $\pm$ 0.0010 & 0.5003 $\pm$ 0.0008 & 0.1040 $\pm$ 0.0020 \\
NS-N2N \cite{Zhou_2024_NSN2N} & \underline{30.62} $\pm$ 0.1957 & \underline{0.8080} $\pm$ 0.0078 & \underline{0.8944} $\pm$ 0.0041 & \underline{0.4406} $\pm$ 0.0260 & \textbf{0.0777} $\pm$ 0.0037 \\
\rowcolor{gray!10}
\textbf{Ours} & \textbf{31.03} $\pm$ \textbf{0.0519} & \textbf{0.8102} $\pm$ \textbf{0.0027} & \textbf{0.9168} $\pm$ \textbf{0.0007} & \textbf{0.4161} $\pm$ \textbf{0.0016} & \underline{0.0788} $\pm$ \textbf{0.0002} \\ 
\bottomrule
\end{tabular}%
}
\end{table}

\subsection{Robustness Analysis Across Random Seeds}
Self-supervised and zero-shot methods can depend on random initialization and
stochastic masking or sampling. We therefore repeat the LIDC-IDRI \cite{Armato_2011_LIDC} evaluation with three seeds (Table~\ref{tab:seed_robustness}). NGPS attains the strongest mean PSNR, SSIM, FSIM, and HFEN among the evaluated methods. Compared with NS-N2N \cite{Zhou_2024_NSN2N}, it also exhibits lower seed variance across all five metrics.

\subsection{Volume-Level Paired Confidence Intervals on LIDC-IDRI}
\label{subsec:supp_lidc_paired_ci}

To assess stability across test subjects, we compute paired metric differences on the six LIDC-IDRI \cite{Armato_2011_LIDC} test volumes. For each volume, the difference is oriented so that a positive value favors NGPS: $\Delta=\mathrm{NGPS}-\mathrm{NS\text{-}N2N}$ for PSNR, SSIM, and FSIM, and $\Delta=\mathrm{NS\text{-}N2N}-\mathrm{NGPS}$ for HFEN and GMSD. The 95\% confidence interval is $\bar{\Delta}\pm t_{0.975,5}s_{\Delta}/\sqrt{6}$.

\begin{table}[t]
\centering
\caption{Volume-level paired differences between NGPS and NS-N2N \cite{Zhou_2024_NSN2N} on the six
LIDC-IDRI \cite{Armato_2011_LIDC} test volumes. Positive $\Delta$ favors NGPS.}
\label{tab:supp_lidc_paired_ci}
\begin{tabular}{lcc}
\toprule
\textbf{Metric} & \textbf{Mean $\Delta$} & \textbf{95\% CI} \\
\midrule
PSNR $\uparrow$ & +0.4100 & [0.2257, 0.5943] \\
SSIM $\uparrow$ & +0.0022 & [-0.0050, 0.0094] \\
FSIM $\uparrow$ & +0.0224 & [0.0184, 0.0264] \\
HFEN $\downarrow$ & +0.0245 & [-0.0020, 0.0510] \\
GMSD $\downarrow$ & -0.0011 & [-0.0049, 0.0027] \\
\bottomrule
\end{tabular}

\end{table}

The intervals exclude zero for PSNR and FSIM. The SSIM, HFEN, and GMSD intervals include zero; the mean GMSD marginally favors NS-N2N \cite{Zhou_2024_NSN2N}. We therefore interpret the LIDC-IDRI results as stable PSNR/FSIM gains and a positive mean HFEN trend rather than uniform dominance across all metrics.

\section{Additional Qualitative Results}
\label{sec:supp_qualitative}

\subsection{Visual Comparisons on the IXI Dataset}
\begin{figure}[htpb]
\centering
\begin{subfigure}{\textwidth}
    \centering
    \includegraphics[width=\textwidth]{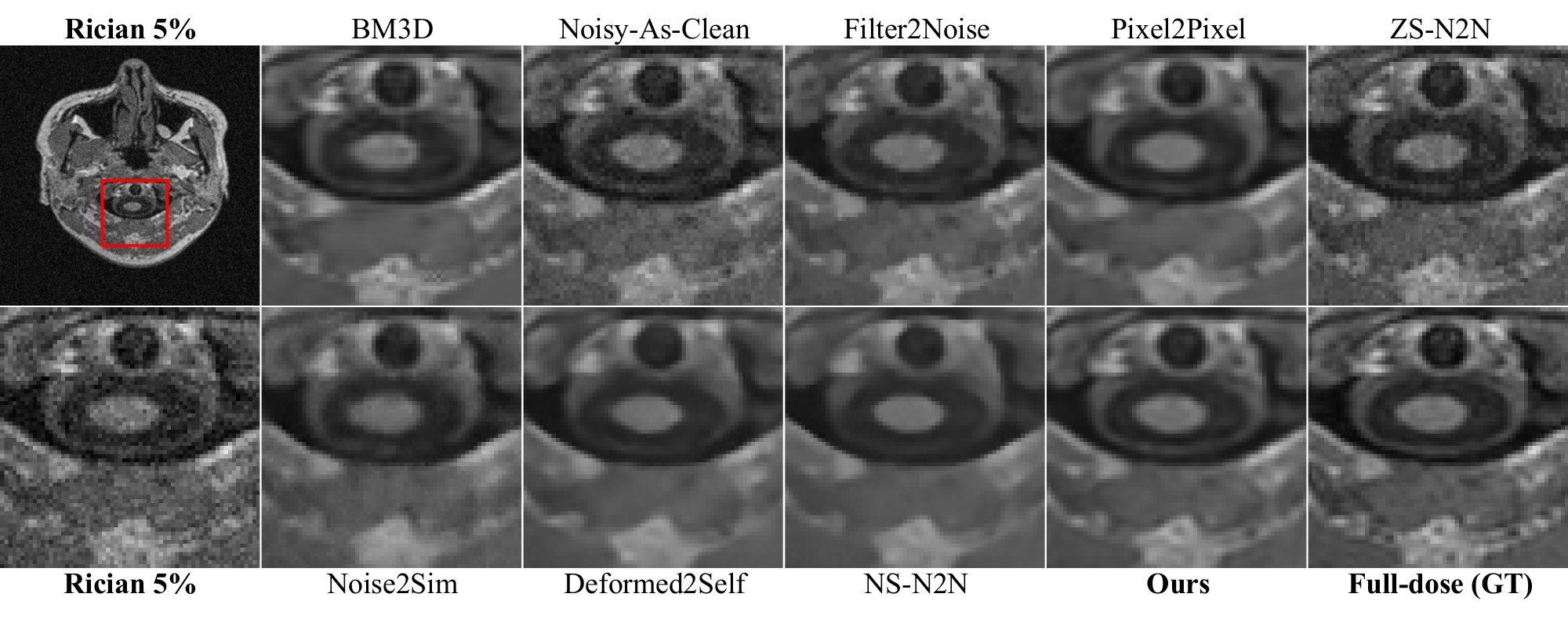}
    \caption{5\% Rician Noise}
    \label{fig:rician5}
\end{subfigure}

\vspace{0.5cm} 

\begin{subfigure}{\textwidth}
    \centering
    \includegraphics[width=\textwidth]{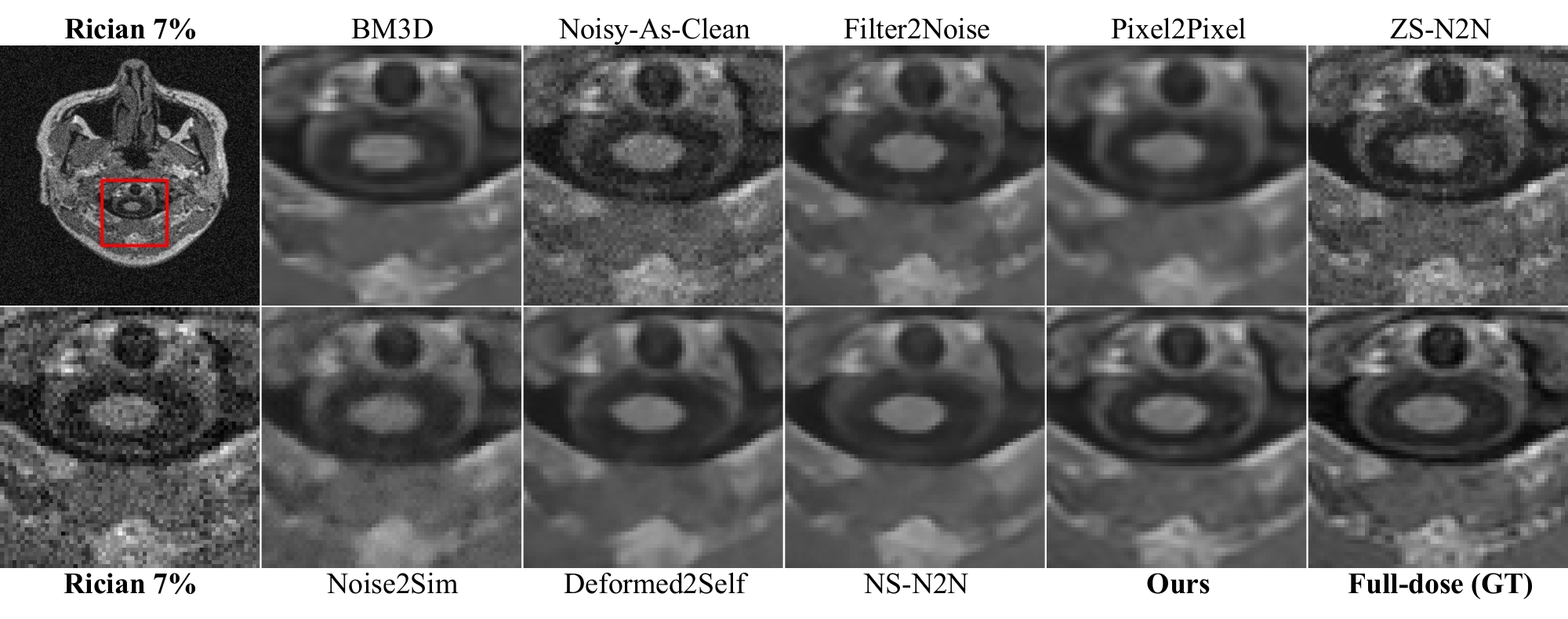}
    \caption{7\% Rician Noise}
    \label{fig:rician7}
\end{subfigure}

\vspace{0.5cm}

\begin{subfigure}{\textwidth}
    \centering
    \includegraphics[width=\textwidth]{figures/fig_rician9.pdf}
    \caption{9\% Rician Noise}
    \label{fig:rician9}
\end{subfigure}
\caption{\textbf{Qualitative comparisons on IXI
\cite{Biomedical_2018_IXI} under 5\%, 7\%, and 9\% synthetic Rician noise.}
At stronger corruption levels, several baselines retain more residual corruption or show softer boundaries in the highlighted regions, whereas NGPS retains sharper local structures in these examples.}
\label{fig:supp_ixi_qualitative}
\end{figure}

Figure~\ref{fig:supp_ixi_qualitative} compares the tested methods under 5\%, 7\%, and 9\% synthetic Rician corruption. At the stronger corruption levels, several baselines retain more residual noise or produce softer boundaries in the highlighted regions. NGPS retains sharper local structures in these examples. These observations are interpreted together with the quantitative results.

\section{Ablation Study}
\label{sec:ablation}

This section evaluates the loss components, PixelBank-style \cite{Ma_2025_Pixel2Pixel} counterfactuals, guide construction, spatial hyperparameters, and controlled failure regime of NGPS.

\subsection{Loss-Component Ablation}

Table~\ref{tab:ablation} compares reconstruction, regional-consistency ($\mathcal{L}_{RC}$), and inter-slice-continuity ($\mathcal{L}_{IC}$) terms on AAPM \cite{McCollough_2017_AAPM}. For NS-N2N \cite{Zhou_2024_NSN2N}, $\mathcal{L}_{IC}$ provides a surrogate constraint in regions omitted by the masked reconstruction loss. For NGPS, which supplies retrieved targets in these regions, adding $\mathcal{L}_{IC}$ is associated with a 0.10 dB decrease.

\begin{table}[t]
\centering
\caption{\textbf{Loss-component ablation on AAPM-Mayo \cite{McCollough_2017_AAPM}.} We compare reconstruction, regional consistency (RC), and inter-slice continuity (IC). IC supplies a surrogate constraint for masked NS-N2N regions, whereas adding it to NGPS is associated with a 0.10 dB decrease.}

\label{tab:ablation}
\renewcommand{\arraystretch}{1.2}
\begin{tabular}{@{}l | ccc | cc@{}}
\toprule
\textbf{Method} & \textbf{Recon.} & \textbf{RC} ($\lambda=0.5$) & \textbf{IC} ($\lambda=1.0$) & \textbf{PSNR (dB)} $\uparrow$ & \textbf{SSIM} $\uparrow$ \\ \midrule
\multirow{4}{*}{NS-N2N \cite{Zhou_2024_NSN2N}}
& \checkmark & & & 35.43 & 0.8481 \\
& \checkmark & \checkmark & & 35.88 & 0.8578 \\
& \checkmark & & \checkmark & 35.44 & 0.8473 \\
& \checkmark & \checkmark & \checkmark & \textbf{35.91} & \textbf{0.8584} \\ \midrule
\multirow{4}{*}{\textbf{NGPS (Ours)}}
& \checkmark & & & 36.50 & 0.8894 \\
& \checkmark & \checkmark & & \textbf{36.68} & \textbf{0.8986} \\
& \checkmark & & \checkmark & 36.39 & 0.8938 \\
& \checkmark & \checkmark & \checkmark & 36.58 & 0.8981 \\ \bottomrule
\end{tabular}
\end{table}

\subsection{PixelBank-Style Counterfactual}
\label{subsec:supp_PixelBank_counterfactual}

\begin{table}[t]
\centering
\caption{PixelBank-style counterfactual training results. Entries are PSNR / SSIM.}
\label{tab:supp_PixelBank_counterfactual}
\resizebox{0.82\textwidth}{!}{%
\begin{tabular}{lccc}
\toprule
\textbf{Method} & \textbf{AAPM} &
\textbf{LIDC 1.25 mm} & \textbf{LIDC 2.5 mm} \\
\midrule
Same-slice PixelBank
& 35.16 / 0.842 & 30.23 / 0.786 & 29.94 / 0.733 \\
Adjacent-slice raw-patch bank
& 36.09 / 0.860 & 30.67 / 0.792 & 30.18 / 0.760 \\
\rowcolor{gray!10}
\textbf{NGPS}
& \textbf{36.68 / 0.899} &
\textbf{31.08 / 0.810} &
\textbf{30.92 / 0.799} \\
\bottomrule
\end{tabular}%
}

\end{table}

Table~\ref{tab:supp_PixelBank_counterfactual} compares NGPS with two simplified PixelBank-style \cite{Ma_2025_Pixel2Pixel} target constructions. The same-slice variant tests intra-image target retrieval, whereas the adjacent-slice raw-patch bank tests the direct extension of raw patch search across slices.

NGPS outperforms the adjacent-slice raw-patch bank by 0.41--0.74\,dB and also outperforms the same-slice variant on all three settings. These comparisons show that neither simplified alternative reproduces the performance of the full NGPS pipeline; they do not isolate the effect of the discrepancy mask or attribute the margin to any single component.

\subsection{Why NS-N2N Uses $\mathcal{L}_{IC}$}
In NS-N2N \cite{Zhou_2024_NSN2N}, misaligned regions are omitted from the masked reconstruction loss.
The inter-slice continuity term therefore supplies a surrogate constraint by encouraging local linearity between averaged inputs and outputs:
\begin{equation}
\mathcal{L}_{IC} = \left\| f_{\theta}\left( \frac{y_z + y_{z+1}}{2} \right) - \frac{f_{\theta}(y_z) + f_{\theta}(y_{z+1})}{2} \right\|_2^2.
\label{eq:lic_def}
\end{equation}
The term follows a first-order Taylor approximation under $f'_{\theta}(y_z)\approx f'_{\theta}(y_{z+1})$ \cite{Zhou_2024_NSN2N}. Adding RC and IC to NS-N2N improves PSNR from 35.43 to 35.91 dB in Table~\ref{tab:ablation}.

\subsection{Interaction Between $\mathcal{L}_{IC}$ and NGPS}
NGPS supplies retrieved targets in the regions omitted by masking. Its best tested configuration uses reconstruction and RC without IC (36.68 dB). Adding IC yields 36.58 dB, indicating empirical objective tension between the retrieved-target reconstruction term and the local-linearity regularizer. Accordingly, the final NGPS objective uses RC but omits IC.

\subsection{Impact of Guide Generation Filters}
\label{subsec:guide_ablation}

\begin{table}[t]
\centering
\caption{\textbf{Ablation study on the guide generation filter.} We evaluate the impact of different filtering strategies on both the AAPM-Mayo and LIDC-IDRI datasets \cite{McCollough_2017_AAPM,Armato_2011_LIDC}. Time is measured in seconds required to process a triplet of adjacent slices (3 slices) on a CPU. The best results are highlighted in \textbf{bold}, and the second-best are \underline{underlined}.}
\label{tab:filter_ablation}
\resizebox{\textwidth}{!}{%
\begin{tabular}{l | c | cc | cc | c}
\toprule
\multirow{2}{*}{\textbf{Filter Type}} & \multirow{2}{*}{\textbf{Threshold ($\tau$)}} & \multicolumn{2}{c|}{\textbf{AAPM-Mayo}} & \multicolumn{2}{c|}{\textbf{LIDC-IDRI}} & \multirow{2}{*}{\textbf{Time (3 slices, s)} $\downarrow$} \\
\cmidrule(lr){3-4} \cmidrule(lr){5-6}
& & \textbf{PSNR} $\uparrow$ & \textbf{SSIM} $\uparrow$ & \textbf{PSNR} $\uparrow$ & \textbf{SSIM} $\uparrow$ & \\
\midrule
None (Raw Noisy) & 0.05 & 35.27 & 0.8825 & 28.94 & 0.7355 & - \\
Gaussian & 0.05 & 36.59 & 0.8935 & 30.67 & 0.7835 & \textbf{0.0084} \\
Median & 0.05 & 36.43 & 0.8939 & 30.38 & 0.7619 & 0.1050 \\
Bilateral & 0.05 & 36.55 & 0.8962 & 30.97 & 0.8047 & \underline{0.0498} \\
\rowcolor{gray!10}
\textbf{Bilateral + Median (Ours)} & 0.05 & \underline{36.68} & 0.8986 & \underline{31.03} & \underline{0.8102} & 0.1422 \\
\midrule
NLM & 0.015 & 36.62 & \textbf{0.8996} & \underline{31.03} & \textbf{0.8117} & 12.6936 \\
NLM + Median \cite{Zhou_2024_NSN2N} & 0.015 & \textbf{36.74} & \underline{0.8988} & \textbf{31.08} & 0.8097 & 12.7077 \\
\bottomrule
\end{tabular}%
}

\end{table}

Table~\ref{tab:filter_ablation} compares guide filters and CPU guide-generation time for a three-slice triplet. NLM variants use $\tau=0.015$ following NS-N2N \cite{Zhou_2024_NSN2N}, whereas the lightweight filters use $\tau=0.05$ to accommodate their larger residual noise. The threshold sensitivity of NGPS is shown in Fig.~\ref{fig:threshold_ablation} of the main paper.

The BF+MF guide provides a practical quality--efficiency trade-off among the tested filters. Gaussian filtering is faster and competitive, whereas the NLM-based \cite{Buades_2005_NLM} guides obtain similar or marginally higher scores but require substantially more preprocessing time under our implementation.

The triplet timings are standalone CPU measurements of guide filtering and are not directly comparable with the vectorized end-to-end 100-slice target-preparation measurements in Table~\ref{tab:time_overhead}.

\subsection{Sensitivity to Spatial Hyperparameters}
\label{subsec:spatial_hyperparam_ablation}

\begin{figure}[t]
\centering
\begin{subfigure}{\textwidth}
    \centering
    \includegraphics[width=\textwidth]{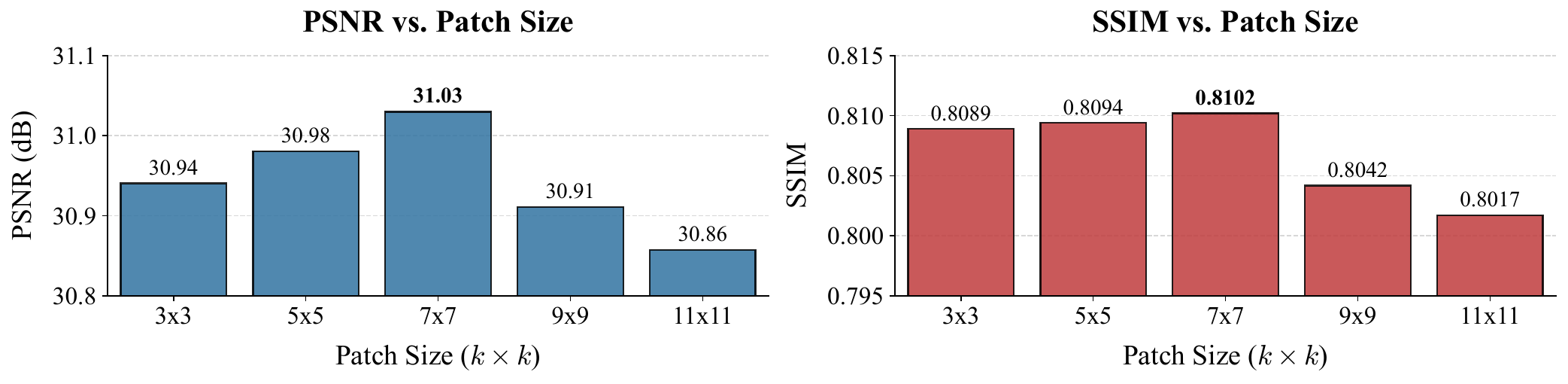}
    \caption{\textbf{Patch-size ablation.} A $7 \times 7$ patch is the best tested setting.}
    \label{fig:patchsize_ablation}
\end{subfigure}
\begin{subfigure}{\textwidth}
    \centering
    \includegraphics[width=\textwidth]{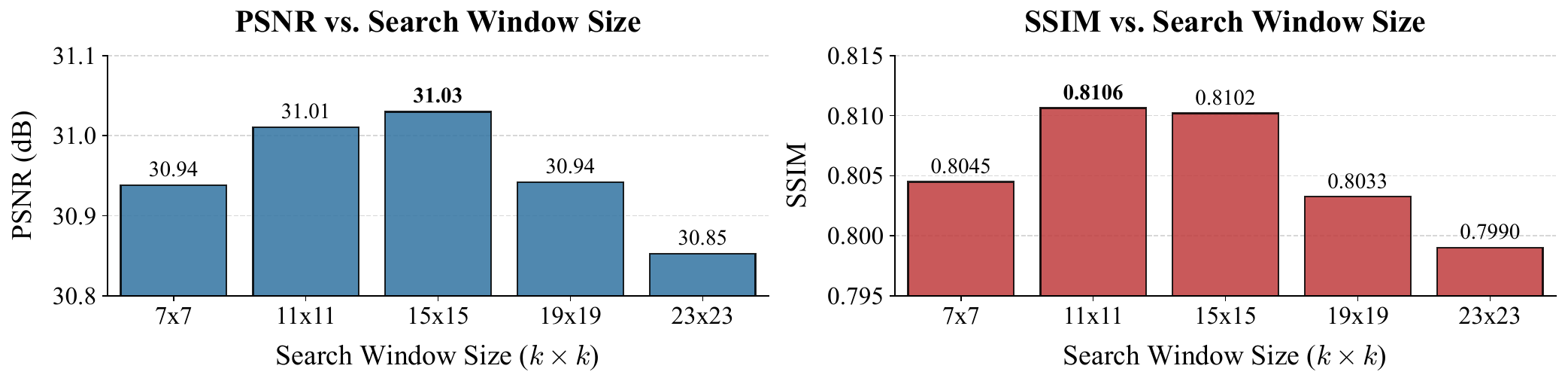}
    \caption{\textbf{Search-window ablation.} A $15 \times 15$ window achieves the highest PSNR in this sweep.}
    \label{fig:windowsize_ablation}
\end{subfigure}

\caption{\textbf{Spatial-hyperparameter ablations on LIDC-IDRI
\cite{Armato_2011_LIDC}.}
Performance varies moderately over the tested patch- and search-window ranges; the selected defaults are marked in the two panels.}

\label{fig:spatial_hyperparam_ablation}
\end{figure}

Recovering displaced supervision depends on matching-patch and search-window sizes. Figure~\ref{fig:spatial_hyperparam_ablation} reports the LIDC-IDRI \cite{Armato_2011_LIDC} sweep. A $7\times7$ patch and $15\times15$ window are the best tested settings; performance remains within 0.18 dB across the evaluated ranges.

\subsubsection{Spacing-Specific Search-Window Sensitivity}
\label{subsec:supp_window_geometry}

To test whether the common search window is specific to thin slices, we vary only $W\in\{7,11,15,19,23\}$ on LIDC-IDRI 1.25 and 2.5 mm while fixing $p=7$, $K=4$, and $\tau=0.05$.

\begin{table}[t]
\centering
\caption{Search-window sensitivity across LIDC-IDRI \cite{Armato_2011_LIDC} slice spacings. Entries are PSNR / SSIM. Relative candidate cost is $W^2/15^2$.}
\label{tab:supp_window_spacing}
\begin{tabular}{cc|cc}
\toprule
$W$ & \textbf{Relative cost} &
\textbf{1.25 mm} & \textbf{2.5 mm} \\
\midrule
7  & 0.22$\times$ & 31.01 / 0.8085 & 30.82 / 0.7938 \\
11 & 0.54$\times$ & \textbf{31.10 / 0.8118} & 30.88 / 0.7965 \\
15 & 1.00$\times$ & 31.08 / 0.8103 & \textbf{30.92 / 0.7994} \\
19 & 1.60$\times$ & 31.02 / 0.8062 & 30.85 / 0.7941 \\
23 & 2.35$\times$ & 30.93 / 0.8015 & 30.76 / 0.7892 \\
\bottomrule
\end{tabular}

\end{table}

The $15\times15$ default is within 0.02 dB of the best 1.25 mm result and is best at 2.5 mm. Larger windows increase candidate cost and reduce performance. We therefore treat $W=15$ as a practical common default over the evaluated geometries, not a universal optimum.

\subsection{Controlled Through-Plane Gap Stress}
\label{sec:supp_gap_stress}

We retrain each method with farther supervisory slices \(z\pm k\), while keeping all hyperparameters fixed to their native-gap settings. The tested gaps are 1--5\,mm for AAPM \cite{McCollough_2017_AAPM} and 1.25--6.25\,mm for the LIDC-IDRI \cite{Armato_2011_LIDC} 1.25\,mm subset.

\paragraph{Optional calibrated match-cost gate (CMG).}
CMG is evaluated only in this stress test and is not part of the base NGPS
method. It suppresses retrieved targets whose guide-patch matching cost is
large relative to costs observed in nominally static regions.
For clarity, we omit the direction index $z\to z'$ from $M$, $F$, $g$, and
$t$ below.

For each location used for calibration or gating, we compute the normalized
mean Top-$K$ matching cost
\begin{equation}
c(p)
=
\frac{1}{Ks^2}
\sum_{j=1}^{K}
\left\|
\mathcal P_s(\tilde y_z,p)
-
\mathcal P_s(\tilde y_{z'},q^{(j)})
\right\|_2^2,
\qquad
s=7,\quad K=4,
\end{equation}
where $q^{(j)}$ denotes the $j$th selected match within the
$15\times15$ search window.

For each slice direction, the calibration set $\mathcal C$ is the first set
containing at least 128 pixels in
\[
\{M=0,F=1\}
\rightarrow
\{M=0\}
\rightarrow
\{F=1\}
\rightarrow
\text{all pixels},
\]
where
\[
M(p)
=
\mathbf 1
\left[
|\tilde y_z(p)-\tilde y_{z'}(p)|>0.05
\right],
\qquad
F(p)
=
\mathbf 1[\tilde y_z(p)>0.01].
\]
The direction-specific threshold is the 95th percentile of the selected
calibration costs:
\[
\gamma
=
Q_{0.95}
\big(
\{c(p):p\in\mathcal C\}
\big),
\qquad
g(p)
=
\mathbf 1[c(p)\leq\gamma].
\]

Rejected targets are not replaced by another pseudo target and receive zero
weight in the dynamic reconstruction term:
\begin{equation}
\mathcal L_{NGPS}^{CMG}
=
\frac{
\sum_p
M(p)g(p)
\big(f_\theta(y_z)(p)-t(p)\big)^2
}{
\sum_p M(p)+\epsilon
}.
\end{equation}
The denominator remains the full flagged-pixel count, rather than the number
of accepted targets. Thus, rejection reduces the contribution of uncertain
dynamic targets instead of re-normalizing the loss over the accepted subset.
The static N2N and regional-consistency terms remain unchanged.

\begin{table}[t]
\centering
\caption{PSNR (dB) under increasing through-plane gap and optional CMG behavior. Bold indicates the best base method in each column.
\(\Delta_{\rm CMG}
=\mathrm{PSNR}_{NGPS+CMG}-\mathrm{PSNR}_{NGPS}\);
rejection is the percentage of flagged targets assigned zero gate weight.}
\label{tab:supp_gap_stress_absolute}
\resizebox{\textwidth}{!}{%
\begin{tabular}{l|ccccc|ccccc}
\toprule
& \multicolumn{5}{c|}{\textbf{AAPM gap (mm)}} &
\multicolumn{5}{c}{\textbf{LIDC 1.25 mm gap (mm)}} \\
\textbf{Method / quantity} & 1 & 2 & 3 & 4 & 5 &
1.25 & 2.5 & 3.75 & 5.0 & 6.25 \\
\midrule
Same-coordinate N2N
& 35.62 & 34.96 & 33.94 & 31.72 & 30.55
& 30.45 & 29.78 & 27.80 & 27.15 & 25.36 \\
Noise2Sim
& 35.49 & 35.38 & 34.96 & 33.82 & 33.27
& 30.53 & 30.22 & 29.72 & 28.30 & 27.42 \\
NS-N2N
& 35.91 & 35.86 & 35.61 & \textbf{35.37} & \textbf{35.12}
& 30.83 & 30.71 & \textbf{30.39} & \textbf{30.28} & \textbf{29.91} \\
\rowcolor{gray!10}
\textbf{NGPS}
& \textbf{36.68} & \textbf{36.51} & \textbf{36.10} & 35.29 & 34.48
& \textbf{31.08} & \textbf{30.96} & 30.34 & 29.52 & 28.68 \\
\midrule
\(\Delta_{\rm CMG}\) (dB)
& -.06 & +.02 & +.12 & +.27 & +.72
& -.06 & -.02 & +.22 & +.60 & +.96 \\
CMG rejection (\%)
& 17.5 & 18.2 & 23.4 & 31.8 & 35.2
& 38.3 & 43.6 & 48.2 & 58.7 & 61.4 \\
\bottomrule
\end{tabular}%
}
\end{table}

Table~\ref{tab:supp_gap_stress_absolute} shows that NGPS leads the base methods at native and small gaps, but falls below NS-N2N \cite{Zhou_2024_NSN2N} at 4--5\,mm on AAPM and 3.75--6.25\,mm on LIDC-IDRI, exposing the fixed-window local-homology limit. CMG is nearly neutral at native gaps but becomes more beneficial as the gap increases, reaching \(+0.72\) and \(+0.96\)\,dB at the largest AAPM and LIDC-IDRI gaps, respectively.

\end{document}